\documentclass[smallextended]{svjour3}       

\usepackage[table,xcdraw,dvipsnames]{xcolor} 
\usepackage[utf8]{inputenc} 
\usepackage[T1]{fontenc}
\usepackage{graphicx}
\usepackage{glossaries}
\usepackage{soul} 
\usepackage[bookmarksdepth=2]{hyperref} 
\usepackage[misc,geometry]{ifsym}  
\usepackage{float} 
\usepackage[sort&compress,square,comma,numbers]{natbib}
\usepackage[many]{tcolorbox} 
\usepackage{enumitem} 
\usepackage{ifthen} 
\usepackage{amssymb} 
\usepackage{booktabs}
\usepackage{graphicx}
\usepackage{makecell}
\usepackage{amsmath}
\usepackage[export]{adjustbox}
\usepackage{ragged2e} 
\usepackage{relsize} 
\usepackage{multirow} 
\usepackage{pbox} 
\usepackage{balance} 
\usepackage{moresize} 
\usepackage{xspace}

\newtcolorbox{mybox}[2][]{
top=0.15in,left=4pt,right=4pt,bottom=4pt,
fonttitle=\bfseries,
colbacktitle=gray,
colback=gray!5,
colframe=gray!40!black,
enhanced,
attach boxed title to top left={xshift=1.5em,yshift=-\tcboxedtitleheight/2},
boxed title style={size=small},
drop shadow={black!50!white},
title=#2,#1}

\newcommand{\countobservations}{
    \def \countobservations{1}
}
\newcounter{observation}
\newenvironment{observation}[1]
{   \refstepcounter{observation}
    \noindent \textbf{Observation~\theobservation) \textit{#1}}
}

\countobservations

\newcommand{\countimplications}{
    \def \countimplications{1}
}
\newcounter{implication}
\countimplications

\newboolean{showcomments}
\setboolean{showcomments}{true}

\ifthenelse{\boolean{showcomments}}
{\newcommand{\nbc}[3]{
 {\colorbox{#3}{\bfseries\sffamily\scriptsize\textcolor{white}{#1}}}
 {\textcolor{#3}{\sf\small$\blacktriangleright$\textit{#2}$\blacktriangleleft$}}
 }
}
{\newcommand{\nbc}[3]{}
 }

\newcommand{\smalltt}[1]{\ifmmode{\mbox{\smaller\texttt{#1}}}\else{\smaller\tt #1}\fi}
\newcommand{\code}[1]{\smalltt{#1}}

\newcolumntype{L}[1]{>{\raggedright\let\newline\\\arraybackslash\hspace{0pt}}m{#1}}
\newcolumntype{C}[1]{>{\centering\let\newline\\\arraybackslash\hspace{0pt}}m{#1}}
\newcolumntype{R}[1]{>{\raggedleft\let\newline\\\arraybackslash\hspace{0pt}}m{#1}}

\usepackage{soul}
\usepackage{moresize}
\usepackage{listings}

\begin{document}

	\title{Predicting long time contributors with knowledge units of programming languages: an empirical study}
	
	\titlerunning{Predicting LTCs with KUs of programming languages: an empirical study}
	
	\author{Md Ahasanuzzaman \and Gustavo A. Oliva \and Ahmed E. Hassan}

	\institute{
		\Letter \space Md Ahasanuzzaman, Gustavo A. Oliva, and Ahmed E. Hassan \at
		Software Analysis and Intelligence Lab (SAIL), School of Computing \\
		Queen's University, Kingston, Ontario, Canada\\    
		\email{\{ma87,gustavo,ahmed\}@cs.queensu.ca}
	}

	\date{Received: date / Accepted: date}

	\maketitle

	\begin{abstract}
		\justifying{
Long-time contributors (LTCs) are essential for the sustainability of open source projects, but unfortunately many  developers leave early. Predicting potential LTCs early in their tenure allows project maintainers to effectively allocate resources and mentoring to enhance their development and retention. Prior study shows that developers are primarily motivated to join FLOSS projects by opportunities to learn and enhance their skills at different areas, including the aspects of programming languages. This motivation plays a crucial role in their continued engagement and contributions to projects. Mapping programming language expertise to developers and characterizing projects in terms of how they use programming languages can help identify developers who are more likely to become LTCs. However, prior studies on predicting LTCs do not consider programming language skills. Towards filling this gap, this paper reports an empirical study on the usage of knowledge units (KUs) of the Java programming language to predict LTCs. A KU is a cohesive set of key capabilities that are offered by one or more building blocks of a given programming language. We select 75 real-world actively maintained Java projects from GitHub. Next, we build a prediction model called KULTC, which leverages KU-based features along five different dimensions. To engineer these features, we detect and analyze KUs from the studied 75 Java projects (spanning a total of 353K commits and 168K pull requests) as well as 4,219 other Java projects in which the studied developers previously worked (spanning a total of 1.7M commits). We compare the performance of KULTC with the state-of-the-art model, which we call BAOLTC. Even though KULTC focuses exclusively on the programming language perspective, KULTC achieves a median AUC of at least 0.75 and significantly  outperforms BAOLTC. Combining the features of KULTC with the features of BAOLTC results in an enhanced model (KULTC+BAOLTC) that significantly outperforms BAOLTC across different settings with a normalized AUC  improvement of 16.5\%. Our feature importance analysis with SHAP reveals that developer expertise in the studied project is the most influential feature dimension for predicting LTCs. Finally, we develop a cost-effective model (KULTC\_DEV\_EXP+BAOLTC) that significantly outperforms BAOLTC. These encouraging results can be helpful to researchers who wish to further study the developers' engagement/retention to FLOSS projects or build models for predicting LTCs. Future work in this area should thus (i) consider KULTC as a baseline model and (ii) consider KU-based features in the design of models that predict LTCs.
}

	\end{abstract}

    \newcommand\PrelimStudy{Preliminary Study Title}
	
    \newcommand\RQOne{How well can KULTC predict LTCs?}

    \newcommand\RQTwo{What is the most important feature dimension of KULTC for predicting LTCs?}

	\newcommand\RQThree{Can KULTC be made more accurate by combining it with BAOLTC?}

	\newcommand\RQFour{How different classification algorithms with hyper-parameter tuning impact the combined model (KULTC+BAOLTC) for predicting long time contributors?}

	\newcommand\RQFive{How does the performance of a cost-effective combined model compare to that of BAOLTC for predicting LTCs?}

    \section{Introduction}
\label{sec:Intro}



Long time contributors (LTCs) are those developers who stay in a project for a long time and contribute a large proportion of code, in turn providing greater value to the community than others, and playing a critical role in  the long-term sustainability of the community~\citep{zhou2010developer,dinh2005freebsd, goeminne2011evidence, mockus2002two}. Unfortunately, many developers in a Free-Libre open source software (FLOSS) projects tend to leave early~\citep{foucault2015impact}. Identifying developers who are likely to become LTCs during their early (e.g, during the first month) participation in a project is important for maintaining the long-term viability of a project. By predicting the most likely future LTCs, project maintainers can efficiently allocate their limited resources, such as directing focused mentoring efforts towards these promising candidates to foster their development and retention~\citep{steinmacher2021being, balali2020recommending}. The success of a FLOSS project is significantly higher when it manages to attract skilled developers and retain them.

A key motivation for developers to contribute to FLOSS projects is the desire to acquire knowledge and enhance their skills. This enduring incentive was highlighted by \citet{gerosa2021shifting}, which confirms that learning and improving skills remain the top-ranked motivation for developers. This finding is also consistent with research conducted two decades earlier~\citep{gerosa2021shifting,lakhani2003hackers}. Developers are particularly driven to contribute by the desire to advance their expertise in various areas, including in programming languages~\citep{silva2020google}. Developers might be more inclined to remain engaged with projects that offer them opportunities to either learn or enhance their current expertise in different aspects of a programming language (e.g., the language's native APIs for concurrency). Hence, we conjecture that mapping programming language expertise to developers and characterizing projects in terms of how they use programming languages can help identify developers who are more likely to become LTCs. However, prior studies on predicting LTCs do not consider programming language skills. Instead, they focus mostly on development activity traces (e.g., number of submitted issues, number of submitted commit comments, and number of submitted pull requests~\citep{zhou2014will, eluri2021predicting}) and other collaboration network features (e.g., degree centrality~\citep{bao2019large}). 

Towards filling this gap, this paper reports an empirical study on the usage of \textit{knowledge units} (KUs) of the Java programming language to predict LTCs. As defined in our prior study~\citep{ahasanuzzaman2024using}, a KU is a cohesive set of key capabilities that are offered by one or more building blocks of a given programming language (see Figure~\ref{fig:ku_metamodel}). For instance, some key capabilities offered by the \textit{concurrency} building block of the Java programming language include (i) creating worker threads to concurrently execute tasks and (ii) using \texttt{synchronized} keyword and \texttt{java.util.concurrent.atomic} package to control the order of thread execution. 

We proceed as follows. First, we select 75 real-world actively maintained Java projects from GitHub. Next, we build a prediction model called KULTC, which uses KU-based features for predicting LTCs in the studied projects. The core idea of the KULTC model is to represent expertise of developers and characteristics of projects using KUs. To this end, we detect and analyze KUs from the studied 75 Java projects (spanning a total of 353K commits and 168K pull requests) as well as 4,219 other Java projects in which studied developers previously worked on (spanning a total of 1.7M commits) before making their initial commits to the studied projects (i.e., joining the studied projects). 
Finally, we engineer our KU-based features along five different dimensions and use them to build our KULTC model. Three of these dimensions are related to expertise: (i) developer expertise in studied projects (KULTC\_DEV\_EXP), (ii) developer expertise in previous projects (KULTC\_PREV\_EXP) and (iii) collaborator expertise in studied projects (KULTC\_COLLAB\_EXP). The other two dimensions focus on project characteristics: (iv) the characteristics of the studied projects (KULTC\_PROJ) and (v) the characteristics of previous projects (KULTC\_PREV\_PROJ). In terms of machine learning algorithm, we use a random forest classifier with default parameter settings of the scikit-learn python package (e.g., number of tree = 100), which is found to be effective based on evidence from prior work~\citep{bao2019large}.

Similar to prior studies~\citep{bao2019large,zhou2014will}, we classify developers as LTCs if they (i) stay in a project for more than a certain time \textit{T} and (ii) commit more than 10\% of the other developers in each year inside \textit{T}. Similar to the study of \citet{bao2019large}, we pick three different settings for \textit{T}: 1 year (LTC-1), 2 years (LTC-2) and 3 years (LTC-3). We thus investigate the effectiveness of KULTC in predicting LTCs under settings LTC-1, LTC-2, and LTC-3. More specifically, we address the following five research questions:

\smallskip \noindent \textbf{RQ1: \RQOne} We compare the performance of KULTC to the state-of-the-art LTC prediction model proposed by \citet{bao2019large}. We refer to this model as BAOLTC from now on. BAOLTC is constructed using 63 different features. We extract these 63 features of BAOLTC from the GHTorrent dataset for the studied developers and projects and construct BAOLTC. To compare the performance of KULTC and BAOLTC, we perform a statistical test using a two-sided Wilcoxon signed-rank test. Additionally, we employ Cliff’s delta~\citep{cliff_delta} as an effect size measure to gauge the practical difference between the distributions. Our results indicate that:

\smallskip \noindent \textit{KULTC achieves a median AUC of at least 0.75 across all three settings for predicting LTCs. An AUC value of 0.70 or higher for a prediction model is deemed decent and suitable for use in practical application~\citep{lessmann2008benchmarking,nam2015clami,jiarpakdee2019impact}. Even though KULTC focuses exclusively on the programming language perspective, KULTC significantly outperforms BAOLTC for LTC-1 and LTC-2 with a median effect size. The normalized AUC improvement of KULTC model over BAOLTC is 6.8\% for LTC-1 and 9.1\% for LTC-2. For LTC-3, the effect size is negligible indicating that both models perform similarly.}

\smallskip \noindent \textbf{RQ2: \RQTwo} To determine the feature dimensions that significantly impact KULTC's performance, we utilize SHAP (SHapley Additive exPlanation). SHAP is a robust, flexible, and widely used method for model interpretation~\citep{molnar2020interpretable,Pacheco23,Gopi22}. Our results indicate that:

\sloppy \smallskip \noindent \textit{The KU-based feature dimension representing a developer’s expertise during his first month’s development activity (KULTC\_DEV\_EXP) is the most influential feature dimension of KULTC for predicting LTCs.}

\sloppy \smallskip \noindent \textbf{RQ3: \RQThree} We combine the features of BAOLTC with the KU-based features of KULTC model to determine whether we can obtain a higher-performing model. More specifically, we use all features of KULTC and BAOLTC to build a combined model, which we call KULTC+BAOLTC. Then, we apply the Scott-Knott Effect Size Difference technique (SK-ESD)~\citep{ghotra_feature_importance_ICSME, mittas_ranking_feature_TSE, kla_model_validation} to compare the performance of KULTC+BAOLTC with that of KULTC and BAOLTC. To investigate the feature dimension of KULTC+BAOLTC that most significantly influence predictions, we employ the same SHAP method from RQ2. Our results indicate that:

\sloppy \smallskip \noindent \textit{KULTC+BAOLTC outperforms both KULTC and BAOLTC across all settings for predicting LTCs. The normalized AUC improvement of KULTC+BAOLTC over BAOLTC is 15.6\% for LTC-1, 16.5\% for LTC-2 and 14.5\% for LTC-3. Similar to the feature importance result in RQ2, the KU-based developer expertise in the studied projects is the most influential feature dimension of KULTC+BAOLTC.}

\smallskip \noindent \textbf{RQ4: \RQFour}
\sloppy We build KULTC+BAOLTC using a random forest classifier with default hyper-parameter values (scikit-learn python implementation). Since every classifier has its own strengths and weaknesses that suit different types of data and problems, the optimal performance of a prediction model can be achieved by choosing an appropriate classifier and tuning its hyper-parameters. To explore how different classifiers affect the KULTC+BAOLTC model, we evaluate a mix of traditional and advanced classifiers: k-nearest neighbor (KNN), Naive Bayes (NB), Decision Tree (DT), Random Forest (RF), XGBoost (XGB), and LightGBM (LGBM). We build the KULTC+BAOLTC models using each of these selected classifiers. To determine the most effective model hyper-parameter configuration, we utilize Scikit-learn's GridSearchCV\footnote{\url{https://scikit-learn.org/stable/modules/generated/sklearn.model_selection.GridSearchCV.html}}. We compare the performance and rank different KULTC+BAOLTC models that are built with the selected classifiers using SK-ESD technique~\citep{ghotra_feature_importance_ICSME, mittas_ranking_feature_TSE, kla_model_validation}. Our results indicate that:

\smallskip \noindent \textit{Among the studied classifiers, the KULTC+BAOLTC model that is built using a random forest classifier with default hyper-parameter values ranks as the highest performing model. Hence, we cannot further improve the performance of KULTC+BAOLTC from RQ3 by leveraging other classifiers and hyper-parameter tuning.}

\smallskip \noindent \textbf{RQ5: \RQFive}
Developing predictive models with a broad range of features typically yields high performance, but such a process can be costly and complex due to the need for extensive data collection and processing. A cost-effective model that uses fewer features yet maintains decent performance is thus valuable, as it reduces operational burdens associated with extensive feature engineering. Towards this goal, we build a cost-effective combined model, which we call KULTC\_DEV\_EXP+BAOLTC. This model combines all the features of BAOLTC with only one feature dimension of KULTC (the developer expertise in the studied projects). Our results indicate that:

\sloppy \smallskip \noindent \textit{ The KULTC\_DEV\_EXP+BAOLTC outperforms BAOLTC model across all settings for predicting LTCs. The normalized AUC improvement of the KULTC\_DEV\_EXP+BAOLTC over BAOLTC is 5.3\% for LTC-1 , 6.1\% for LTC-2 and 3.3\% for LTC-3. Therefore, we improve over the state-of-the-art while still keeping feature engineering cost at a reasonable level.}

\medskip \noindent \textbf{The main contribution of our paper are as follows:} (i) introduction of a feature engineering approach tailored to transform KUs of programming languages into KU-based features for predicting LTCs, (ii) the design of a KU-based LTC prediction model (KULTC) (and variations, such as KULTC+BAOLTC and KULTC\_DEV\_EXP+BAOLTC) that are overall superior to the state-of-the-art model for predicting LTCs. Future work in this area should thus consider KU-based prediction models as baselines, as well as  experiment with KU-based features. A supplementary material package is provided online\footnote{\url{https://shorturl.at/dpKU4}}.


\smallskip \noindent \textbf{Paper organization.} Section~\ref{sec:Knowledge_Unit} defines KUs and presents our approach to detect them. Section~\ref{sec:methodology} describes our study methodology, including the data collection, feature engineering approach, and the model construction. Section~\ref{sec:Main_study} presents the motivation, approach, and findings of our three research questions. Section~\ref{sec:Discussion} discusses the trade-off between model complexity vs performance and direction to improvement of models. Section~\ref{sec:Related_Work} discusses related work. Section~\ref {sec:Limitations_And_Threats} describes the threats to the validity of our findings. Finally, Section~\ref{sec:Conclusion} outlines our concluding remarks.
	\section{Knowledge Units (KUs)}
\label{sec:Knowledge_Unit}

We propose knowledge units (KUs) of programming languages in our prior work~\citep{ahasanuzzaman2024using}. To ensure that this paper is as self-contained as possible, in the following we briefly introduce KUs (Section~\ref{subsec:knowledge_definition}) and discuss our approaches for eliciting (Section~\ref{subsec:knowledge_operational_definition}) and detecting them (Section~\ref{subsec:knowledge_detection}). 

\subsection{Definition}
\label{subsec:knowledge_definition}

Every programming language consists of \textit{building blocks} that developers use to write code. 
For the Java programming language, these building blocks encompass both the Java fundamental language constructs (e.g., if/else statements, try/catch statements, and synchronize blocks) and APIs (e.g., Concurrency API, Stream API, and the Generic and Collections API). Each building block offers a \textit{set of capabilities} essentially representing the actions a developer can perform using that building block.

\begin{figure}[!t]
    \centering
    \includegraphics[width=1.0\textwidth]{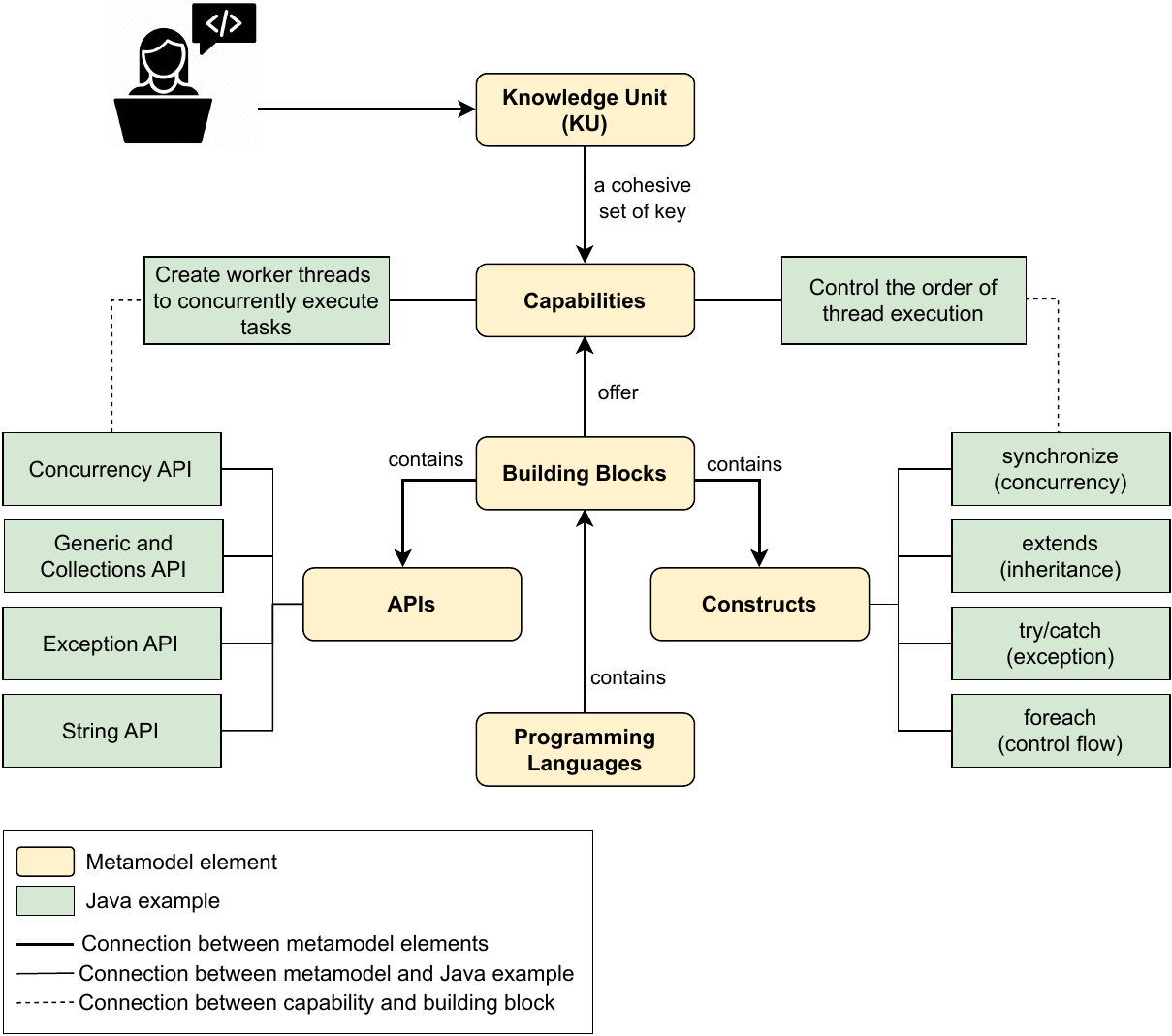}
    \caption{Our metamodel for knowledge units (KUs).}
    \label{fig:ku_metamodel}
\end{figure}

We define a Knowledge Unit (KU) as a cohesive set of key capabilities that are offered by one or more building blocks of a given programming language. The inclusion of ``key'' in our definition aims to ensure that KUs are centered around fundamental capabilities rather than those that are overly specific. Figure~\ref{fig:ku_metamodel} illustrates key capabilities that are associated with the Concurrency API (left-hand side) and the \texttt{synchronize} language construct (right-hand side).

\subsection{Eliciting KUs}
\label{subsec:knowledge_operational_definition}


Programming language certification exams, such as the Oracle Java SE and Java EE certification exams for Java, are designed to assess a developer's expertise and proficiency in using the key  capabilities provided by the building blocks of the language. Therefore, we can say that these certification exams reflects the KUs of a programming language. For example, the Java SE 8 Programmer II certification exam includes a topic on ``Generics and Collections.'' The subtopics of of ``Generics and Collections.'' that are covered in the exam include: (i) create and use a generic class, (ii) create and use \texttt{ArrayList}, \texttt{TreeSet}, \texttt{TreeMap}, and \texttt{ArrayDeque} objects, (iii) use \texttt{java.util.Comparator} and \texttt{java.lang.Comparable} interfaces, and (iv) iterate using \texttt{forEach} methods of a \texttt{List}. We interpret such a list of subtopics as the key capabilities that are offered by the ``Generics and Collections'' building block of the Java programming language. From this interpretation, we infer that Java has a ``Generics and Collections KU''.

\begin{table}[!htbp]
    \centering
    \caption{The identified Java programming language knowledge units (KUs).}
    \label{tab:topic_definition}
    \resizebox{\columnwidth}{!}{
    \begin{tabular}{p{3.2cm}p{13.8cm}}
        \toprule
        \multicolumn{1}{C{3cm}}{\textbf{Knowledge unit (KU)}} & \multicolumn{1}{C{14cm}}{\textbf{Definition}}                                                                                                                                                                                                                               \\ \midrule

        \textbf{[K1]} Data Type                                         &
        The declaration and initialization of different types of variables (e.g., primitive type and parameterized type.)
        \\ \midrule

        \textbf{[K2]} Operator and Decision                             &
        The usage of different Java operators (e.g., assignment, logical, and bit-wise operators) and conditional statements (e.g, if, if-else, and switch statements).
        \\ \midrule

        \textbf{[K3]} Array                                            &
        The declaration, instantiation, initialization and the usage of one-dimensional and multi-dimensional arrays.
        \\ \midrule

        \textbf{[K4]} Loop                                             &
        The execution of a set of instruction-s/methods repeatedly using for, while, and do-while statements and the skipping and stopping of a repetitive execution of instructions and methods using continue and break statements.
        \\ \midrule

        \textbf{[K5]} Method and Encapsulation                         &
        The creation of methods with parameters, the use of overloaded methods and constructors, the usage of constructor chaining, the creation of methods with variable length arguments, and the usage of different access modifiers. This KU also describes encapsulation mechanisms, such as creating a set and get method for controlling data access, generating immutable classes, and updating object type parameters of a method.
        \\ \midrule

        \textbf{[K6]} Inheritance                                      &
        Developing code with child class and parent class relationship, using polymorphism (e.g., developing code with overridden methods), creating abstract classes and interfaces, and accessing methods and fields of the parent's class.
        \\ \midrule

        \textbf{[K7]} Advanced Class Design                             &
        Developing code that uses the final keyword, creating inner classes including static inner classes, local classes, nested classes, and anonymous inner classes, using enumerated types including methods and constructors in an enum type, and developing code using @override annotator.
        \\ \midrule

        \textbf{[K8]} Generics and Collection                           &
        The creation and usage of generic classes, usage of different types of interfaces (e.g., List Interface, Deque Interface, Map Interface, and Set Interface), and comparison of objects using interfaces (e.g., java.util.Comparator, and java.lang.Comparable).
        \\ \midrule

        \textbf{[K9]} Functional Interface                         &
        The development of code that uses different versions of defined functional interfaces (e.g., primitive, binary, and unary) and user-defined functional interfaces.
        \\ \midrule

        \textbf{[K10]} Stream API                                       &
        The development of code with lambda expressions and Stream APIs. This includes developing code to extract data from an object using peek() and map() methods, searching for data with search methods (e.g., findFirst) of the Stream classes, sorting a collection using Stream API, iterating code with foreach of Stream, and saving results to a collection using the collect method.
        \\ \midrule

        \textbf{[K11]} Exception                                      &
        The creation of try-catch blocks, the usage of multiple catch blocks, the usage of try-with-resources statements, the invocation of methods throwing an exception, and the use of assertion for testing invariants.
        \\ \midrule

        \textbf{[K12]} Date time API                                   & This KU refers to create and manage date-based and time-based events using Instant, Period, Duration, and TemporalUnit, and work with dates and times across timezones and manage changes resulting from daylight savings, including Format date and times values. \\ \midrule

        \textbf{[K13]} IO                                              &
        Reading and writing data from console and files and using basic Java input-output packages (e.g., java.io.package).
        \\ \midrule

        \textbf{[K14]} NIO                                              &
        Interacting files and directories with the new non-blocking input/output API (e.g., using the Path interface to operate on file and directory paths). Performing other file-related operations (e.g., read, delete, copy, move, and managing metadata of a file or directory).
        \\ \midrule

        \textbf{[K15]} String Processing                                &
        The knowledge about searching, parsing, replacing strings using regular expressions, and using string formatting.
        \\ \midrule

        \textbf{[K16]} Concurrency                                      &
        The knowledge about functionalities that are related to thread execution and parallel programming. Some of these functionalities include: creating worker threads using Runnable and Callable classes, using an ExecutorService to concurrently execute tasks, using synchronized keyword and java.util.concurrent.atomic package to control the order of thread execution, using java.util.concurrent  classes and using a parallel fork/join framework.
        \\ \midrule

        \textbf{[K17]} Database                                        &
        The creation of database connection, submitting queries and reading results using the core JDBC API.
        \\ \midrule   

        \textbf{[K18]} Localization                                     &
        The knowledge about reading and setting the locale (Oracle defines a locale as ``a specific geographical, political, or cultural region'') by using the Locale object, and building a resource bundle for each locale, and loading a resource bundle in an application.
        \\ \midrule

        \textbf{[K19]} Java Persistence                                  &
        The usage of object/relational mapping facilities for managing relational data in Java applications. With this knowledge, developers can learn how to map, store, update and retrieve data from relational databases to Java objects and vice versa.                                                              \\ \midrule

        \textbf{[K20]} Enterprise Java Bean                             &
        The knowledge about managing server-side components that encapsulate the business logic of an application.                                                                                                                                                                                                               \\ \midrule

        \textbf{[K21]} Java Message Service API                         &
        The knowledge of how to create, send, receive and read messages using reliable, asynchronous, and loosely coupled communication.                                                                                                                                                                                                   \\ \midrule

        \textbf{[K22]} SOAP Web Service                               &
        Creating and using the Simple Object Access Protocol for sending and receiving requests and responses across the Internet using JAX-WS and JAXB APIs.                                                                                                                                                          \\ \midrule

        \textbf{[K23]} Servlet                                          &
        Handling HTTP requests, parameters, and cookies and how to process them on the server sites with appropriate responses.                                                                                                                                                                                           \\ \midrule

        \textbf{[K24]} Java REST API                                    &
        Creating web services and clients according to the Representational State Transfer architectural pattern using JAX-RS APIs.                                                                                                                                                                                     \\ \midrule

        \textbf{[K25]} Websocket                                         &
        Creating and handling bi-directional, full-duplex, and real-time communication between the server and the web browser.                                                                                                                                                                                         \\ \midrule

        \textbf{[K26]} Java Server Faces                                 &
        The knowledge about how to build UI component-based and event-oriented web interfaces using the standard JavaServer Faces (JSF) APIs.                                                                                                                                                                                                     \\ \midrule

        \textbf{[K27]} Contexts and Dependency Injection (CDI)             &
        Managing the lifecycle of stateful components using domain-specific lifecycle contexts and type-safely inject components (services) into client objects.                                                                                                                                                     \\ \midrule

        \textbf{[K28]} Batch Processing                                    &
        Creating and managing long-running jobs on schedule or on demand for performing on bulk-data, and without manual intervention.                                                                                                                                                                                 \\ \bottomrule
    \end{tabular}
    }
\end{table}

We reuse the KUs that we elicited in our prior work~\citep{ahasanuzzaman2024using}. To elicit those KUs, we relied on the Oracle certification exams for the Java programming language. Oracle offers certification exams for different Java editions, such as Java Standard Edition (Java SE) and Java Enterprise Edition (Java EE). Specifically, for Java SE, there are two levels of certification exams:(i) \textit{Oracle Certified Associate, Java SE 8 Programmer I Certification exam}~\citep{oracle_se_oap}, and (ii) \textit{Oracle Certified Professional, Java SE 8 Programmer II Certification exam }\citep{oracle_se_ocp}. For Java EE, only one version of the exam is offered by Oracle: \textit{Oracle Certified Professional, Java EE Application Developer Certification exam}~\citep{oracle_ee_prof}. We elicited KUs directly from the topics outlined in these exams. In the vast majority of cases, an exam topic was interpreted as a KU (e.g., the Generics and Collections  topic of Java SE 8 Programmer II Certification exam was interpreted as ``Generics and Collections KU'') and its subtopics (e.g., create and use a generic class, create and use \texttt{ArrayList}, \texttt{TreeSet}, \texttt{TreeMap}, and \texttt{ArrayDeque} objects, and use \texttt{java.util.Comparator} and \texttt{java.lang.Comparable} interfaces) were interpreted as the key capabilities of that KU. At the end of this process, we elicited 28 KUs (Table~\ref{tab:topic_definition}). For a more detailed description of this mapping process, please refer to our prior work~\citep{ahasanuzzaman2024using}. 

\subsection{Detection of KUs}
\label{subsec:knowledge_detection}

Analogously to our prior work~\citep{ahasanuzzaman2024using}, we use static analysis to detect KUs from a given Java project. Fist, we collect all the commits of a given project. Next, for each collected commit C, we checkout the code snapshot associated with C and parse all the java source files. To parse Java source files we use the Eclipse JDT framework~\citep{eclipse_jdt}. The JDT framework creates an Abstract Syntax Tree\footnote{The grammar employed by JDT can be seen at \url{https://github.com/eclipse-jdt/eclipse.jdt.core/blob/master/org.eclipse.jdt.core.compiler.batch/grammar/java.g}} (AST) for the source code and provide visitors for each element of this tree such as \textit{names}, \textit{types} (e.g., class, and interfaces), \textit{expressions} (e.g., assignment, and postfix), \textit{statements} (e.g., if, break, and for statements) and \textit{declarations} (e.g., blocks, and methods).

To detect a KU, we focus on the specific capabilities of that unit as outlined in Table~\ref{tab:ku-from-java-exams}. For instance, to detect the presence of the \textit{Inheritance KU} ([K6]), we utilize the JDT framework's visitors to recognize capabilities linked to that unit. These include [K6, C1] where \textit{a superclass refers to subclass}, [K6, C3] which involves \textit{creating override methods}, and [K6 C4] which pertains to \textit{ create ``abstract'' classes and ``abstract'' methods}. By employing type and declaration visitors, for instance, we can ascertain if a class overrides a method of another class ([K6, C3]). 

During our code analysis, we also use the JDT to collect binding information of variables, classes, and methods (e.g., \code{org.eclipse.jdt.core.dom.IMethodBinding} resolves method binding information). In our study, we exclude type bindings that are related to third-party libraries since our primary interest lies in studying the KUs that are associated with the source code written by the developers of the studied projects. 

We count the occurrences of KUs for each Java source file. For instance, one of the capabilities associated with the \textit{Inheritance KU} ([K6]) involves the creation of abstract classes ([K6,C4]).  Therefore, if a declaration of an \texttt{abstract} class is detected in a given Java file \code{X.java} belonging to a project \textit{P}, we increment the counter for \textit{Inheritance KU} that is associated with file \texttt{X.java} by one. In other words, for each file, we generate a vector \textit{V} containing the counts of each KU in that file.
    \section{Methodology}
\label{sec:methodology}

In this section, we outline the methodology used in our study, starting with the data collection process (Section~\ref{data_collection}). Next, we discuss the approach for classifying LTCs (Section~\ref{ltc_classification}). We then discuss how we engineer the features for KULTC (Section~\ref{sec:kultc-feateng}). Finally, we describe how we construct (Section~\ref{kultc_construction}) and evaluate (Section~\ref{model_evaluation}) KULTC.

\subsection{Data Collection}
\label{data_collection}

Our data collection process contains three main steps. First, we retrieve the source code and pull request (PR) data from real-world, active Java software projects. We refer to these projects as our studied projects. Then, we select developers of the studied projects. Finally, we collect commit data of developers' previous projects. Figure~\ref{fig:data_collection} presents an overview of our data collection process. We describe each step in model detail below.

\begin{itemize}[wide = 0pt, label=$\bullet$]
    \item \textit{Step 1) Collect commit data and pull request data of the studied projects:}
        \begin{itemize}[wide = 0pt, itemsep = 3pt, topsep=3pt, listparindent=\parindent]
            \item \textit{Step 1.1) Select Java software projects for our study.} In our study, we select the same 75 Java projects that were analyzed by ~\citet{bao2019large} in their recent research on predicting long time contributors. All these projects are popular (e.g., all projects contain more than 100 stars) and actively maintained. Moreover, we note that these projects are listed in the \textit{dataset of engineered projects} that was made available by~\citet{engineered_project_github_EMSE_2017}. The data collection date is September 3rd, 2023.
            
            \item \textit{Step 1.2) Download the commit history of the selected Java software projects.} To download the commit history of our studied software projects, we follow three steps. First, we clone the repository of each project using the~\texttt{git clone} command. Next, we utilize PyDriller~\citep{pydriller_MSR_2018} to collect all the commit information of a project, such as commit hash id, commit author name, commit author date, commit message and the list of modified Java files in a commit. Finally, to retrieve the source code of a particular commit (snapshot) of a project, we use the command~\texttt{git checkout commit\_hash}, where~\texttt{commit\_hash} is the commit id. Our supplementary material package contains the list of our studied projects along with the links to their corresponding GitHub repositories. 

            \item \textit{Step 1.3) Download the pull request (PR) data of the selected Java projects.} We use the GitHub Rest API\footnote{\url{https://docs.github.com/en/rest/reference}} to download the PR data of the studied projects. For instance, to collect all the PRs of the ElasticSearch project, we send a GET request to the following URL: \url{https://api.github.come/repos/elastic/elasticsearch/pulls?state=all\&per\_page=100}. We also use the GitHub API to collect discussion comments and changed files for every PR.
        \end{itemize}
    \item \textit{Step 2) Select developers for our study:} Developer's previous expertise on other projects that they worked on before contributing to the project under investigation (i.e., the studied projects) could help identify long time contributors. Hence, we examine the previous projects of developers (i.e., those that they pushed commits to). To collect the previous projects of developers, we leverage the GHTorrent dataset~\citep{gousios2013ghtorent}. However, we face a challenge: GHTorrent lists developers by their GitHub account's username (i.e., a unique id of a developer across GitHub), while the commit data contains commit author names of developers. Since developers' GitHub username and commit author name can be different, we need to establish a link between GitHub account and commit author name. To maintain objectivity and eliminate any potential bias in our analysis, we follow a conservative approach and select those developers for whom we can confidently establish a link between their commit author names and GitHub accounts. We briefly discuss the steps that we follow in this approach below. 

    \begin{figure}[!t]
        \centering
        \includegraphics[width=1.0\textwidth]{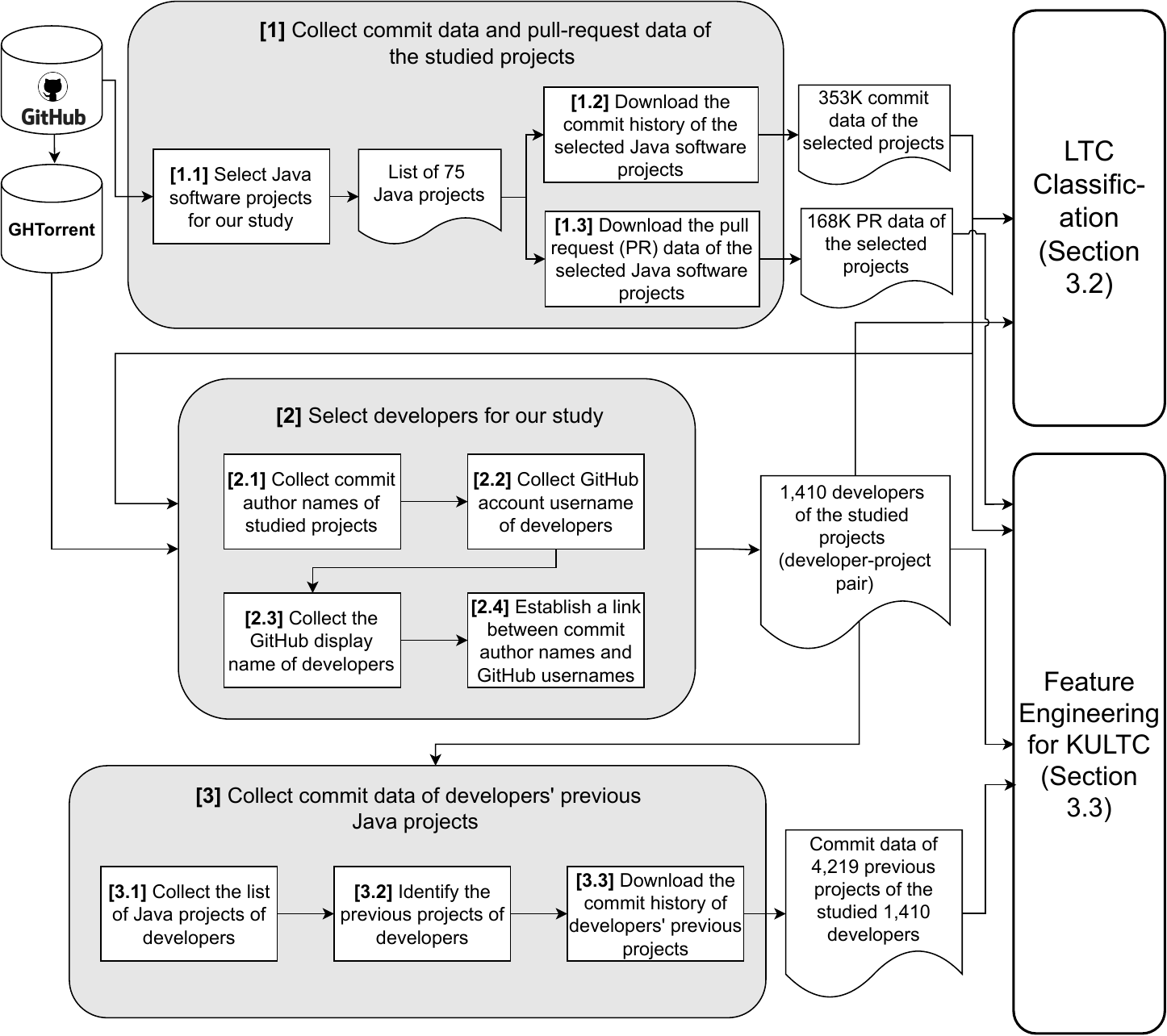}
        \caption{An overview of our data collection process.} 
        \label{fig:data_collection}
    \end{figure}

    
        \begin{itemize}[wide = 0pt, itemsep = 3pt, topsep=3pt, listparindent=\parindent]
            \item \textit{Step 2.1) Collect commit author names of studied projects.} We collect the commit author names of developers of the studied projects from the commit data (c.f., Step 1.2).
            
            \item \textit{Step 2.2) Collect GitHub account username of developers.} We collect the GHTorrent's MySQL database (snapshot of 2021/03/06). For each of the studied projects, we select the developers of the project (i.e., make commit) and collect their GitHub accounts' usernames.
            
            \item \textit{Step 2.3) Collect the GitHub display name of a developer.} GitHub's users can set a display name in their GitHub profile, which can be their real name. To collect the GitHub display name of a developer we send a GET request to the following URL: \url{https://api.github.com/users/username}. Here, \texttt{username} is the GitHub username of a developer. At the end of this step, we collect the GitHub display name of every developer.

            \item \textit{Step 2.4) Establish a link between commit author names and GitHub accounts.} First, for each of the studied projects, we select commit author names of the project. Next, to determine which GitHub user pushed a commit in a project, we search the commit author's name in the list of GitHub display name of the project. We establish a link between the commit author's name to the GitHub display name if the commit author name matches with the GitHub display name. We successfully establish a link for 1,410 developers across the studied projects. We select these 1,410 developers for our study. Finally, we create a developer-project pair (we denote as \textit{Pair(P,D)}) for these 1,410 developers. Here, a project \textit{P} is one of the studied projects and a developer \textit{D} is one of the selected developers of the project \textit{P}.
        \end{itemize}    
    
    \item \textit{Step 3) Collect commit data of developers' previous Java projects:} For each \textit{Pair(P,D)} we look into the contribution history of \textit{D} to find the previous projects that the developer \textit{D} contributed to before his initial in the studied project \textit{P}. The following steps are used to collect commit data of developers' previous Java projects.
        
        \begin{itemize}[wide = 0pt, itemsep = 3pt, topsep=3pt, listparindent=\parindent]
            \item \textit{Step 3.1) Extract the list of Java projects of developers.} For each of the selected developers of every studied project, we first collect the list of all projects they worked on. We extract the list of Java projects where these developers pushed at least one commit using the GHTorrent dataset.

            \item \textit{Step 3.2) Identify the previous Java projects of developers.} For every developer-project pair, \textit{Pair(P,D)}, we first extract the initial commit the developer \textit{D} pushed to the studied project \textit{P} from the commit data. Then, we compare this initial commit's date with the dates of commits \textit{D} pushed to other Java projects (specifically those projects that we selected in the previous step for the developer). We identify a project as a previous project if \textit{D} pushed at least one commit to the project before his initial commit to \textit{P}. At the end of this process, we have a set of previous projects the developers worked on before contributing to the studied projects. That is, we have a set of previous projects for every developer-project pair \textit{Pair(P,D)}. We find that the total number of unique previous Java projects is 4,219.

            \item \textit{Step 3.3) Download the commit history of developers' previous projects.} We follow the same step of Step 1.2 to download the commit history of developer's previous projects. We clone the repository of all 4,219 projects using the \texttt{git clone} command. To collect all the commit information (e.g., commit hash id) of a project, we utilize PyDriller~\citep{pydriller_MSR_2018}. Finally, we retrieve the source code of a specific commit of a project using the \texttt{git checkout commit\_hash}, where commit\_hash is the commit id.
        \end{itemize}
\end{itemize}

\subsection{LTC Classification}
\label{ltc_classification}

To classify a developer as a \textit{long-term contributor} (LTC), we adopt the same criteria that is used in previous studies \citep{bao2019large,zhou2014will}. More specifically, a developer can be an LTC if he simultaneously satisfies the following two conditions:
    \begin{itemize}[itemsep = 3pt, topsep = 3pt]
        \item [(i)]\textit{Duration:} The developer stays with a project for more than a certain time \textit{T}. In other words, the time interval between the first commit and the last commit of a developer in the project of interest is larger than \textit{T}.
        \item [(ii)]\textit{Productivity:} The number of commits of the developer is higher than the 10th percentile across other developers in each year of T.
    \end{itemize} 
Following the approach of \citet{bao2019large}, we use three different settings for T: 1 year (LTC-2), 2 years (LTC-2), and 3 years (LTC-3). We thus investigate the effectiveness of models in predicting LTCs under settings LTC-1, LTC-2, and LTC-3. For each interval T, we apply the LTC criteria on the commit data of the studied projects. The developers who satisfy the LTC criteria are labeled as LTCs and those who do not meet the criteria are labeled as Non-LTCs. 
        
\subsection{Feature Engineering for KULTC}
\label{sec:kultc-feateng}

Our KUTLC model includes features from five distinct dimensions, namely: (i) developer expertise in the studied projects, (ii) developer expertise in previous projects (i.e., developer prior expertise), (iii) collaborator expertise in the studied projects, (iv) characteristics of studied projects, and (v) characteristics of previous projects. We engineer 28 KU-based features in each dimension, yielding a total of 140 KU-based features. In the following, we explain how we engineer those KU-based features for our KULTC model.

\begin{itemize}[wide = 0pt, itemsep = 3pt, topsep=1pt, label=$\bullet$]
    \item \textit{Developer expertise in studied projects (KULTC\_DEV\_EXP).} This dimension refers to the KU-based expertise features of developers in the studied projects. Our assumption is that developers programming language expertise that they demonstrate early in their project contributions could indicate whether they are likely to become long time contributors (or not). Each feature represents the developer's expertise in one of the specific KUs that have been identified in Table~\ref{tab:topic_definition}. For each studied developer-project pair, we first select the developer's initial commit in the project. Then, we create a 30-day window from this initial commit and select all commits pushed by the developer within this 30-day window. Using our custom KU detector built atop Eclipse JDT, we identify and count the occurrences of KUs in all the modified Java files (c.f., Section \ref{subsec:knowledge_detection}). Finally, we sum the count of the occurrences of every KU to quantify the expertise of a developer gained in his/her first month of commits.
    
    
    \item \textit{Developer expertise in previous projects (KULTC\_PREV\_EXP).} This dimension comprises the previous expertise of a developer based on KUs. Each feature reflects a developer's expertise in one of the specific KUs listed in Table~\ref{tab:topic_definition}, which have been acquired through their commit activities in previous projects. For each studied developer-project pair, we analyze the developer's commits from previous projects that are pushed prior to his/her initial commit of the studied project. We detect KUs in changed Java files of these commits and count the sum of the occurrences for every KU. Finally, we calculate the median of these occurrences for every KU across all previous projects to quantify the expertise gained in previous projects.
    

    \item \textit{Collaborator expertise in the studied projects (KULTC\_COLLAB\_EXP).} This dimension refers to the KU-based expertise features of the collaborators of the studied developers in the projects being studied. We hypothesize that expert collaborators can mentor and inspire newcomers, encouraging them to continue working on a project for a long period. Each feature represents the expertise of a developer's collaborators in one of the specific KUs that are outlined in Table~\ref{tab:topic_definition}. To identify a developer's collaborators in a project, we first analyze the discussion comments on PRs submitted by the developer within a 30-day window from their initial commit in the project. We record the authors of these comments as the collaborators. For each collaborator, we analyze the commits that they made prior to the developer’s initial commit, detect KUs in changed Java files, and count their occurrences. We then aggregate these counts for each KU across all files and collaborators. Finally, we calculate the median value of these aggregated counts for each KU to quantify the expertise of collaborators gained in the project.

    \item \textit{Characteristics of the studied projects (KULTC\_PROJ)).} This dimension focuses on the KU-based features that describe the characteristics of the projects being studied. Specifically, it captures the state of the project at the moment a new developer joins, providing a snapshot of the project's attributes at that time. We hypothesize that developers are often drawn to projects that focus on specific aspects of programming languages and are likely to continue working on those projects that align with their interests. For every studied developer-project pair, we select the commit that is pushed immediately prior to the developer's initial commit. We detect KUs in every Java files of the selected commit and count their occurrences. Finally, we sum the count of the occurrences of every KU to depict the project’s characteristic at the point of the new developer's entry.

    \item \textit{Characteristics of previous projects (KULTC\_PREV\_PROJ).} This dimension refers to the KU-based features that represent the characteristics of the previous projects that a developer worked before joining the studied project. It is based on the hypothesis that developers tend to continue working on new projects that involve similar programming aspects to those they have previously focused on. For each previous project of every studied developer-project pair, we select the commit that is pushed immediately prior to the developer's initial commit in the studied project. We identify KUs in all Java files from these commits, count their occurrences, and sum these counts for each KU. Finally, we calculate the median value of these sums for every KU across all previous projects.
    
\end{itemize}

\subsection{KULTC Model Construction}
\label{kultc_construction}

In this section, we describe our KULTC model construction steps. We construct the KULTC model using KU-based features data (c.f., Section~\ref{sec:kultc-feateng}) collected from 1,410 developers across our 75 studied projects. The model's goal is to predict which of these developers will become LTCs in the projects that we are studying. Our approach for constructing the KULTC model is summarized in Figure~\ref{fig:model-construction}. We explain each step in more detail below.

\begin{itemize}[wide = 0pt, itemsep = 3pt, topsep=3pt, listparindent=\parindent, label=$\bullet$]
    
    \item \textit{Step 1) Select non-correlated features.} To better interpret the results of a prediction model, it is recommended to mitigate correlated features~\citep{jiarpakdee2019impact,Chak_ICSE_SEIP}. Therefore, we want to check the performance of our studied models using non-correlated features. We select non-correlated features using the AutoSpearman approach introduced by~\citet{jiarpakdee2019impact}, which is available in the \texttt{Rnalytica} package\footnote{https://github.com/awsm-research/Rnalytica}. AutoSpearman automatically selects non-correlated features based on two analyses: (1) a Spearman rank correlation test and (2) a Variance Inflation Factor analysis. We apply the AutoSpearman on KU features, which resulted in 83 non-correlated features for LTC-1, LTC-2 and LTC-3.

    \item \textit{Step 2) Generate bootstrap samples.} To estimate the performance of our models in practice, we apply the \textit{out-of-sample bootstrap} model validation technique with 100 repetitions~\citep{kla_model_validation, efron1983estimating}. More specifically, we train our models on the 100 bootstrap samples and test their performance using the data that do not appear in the samples (i.e., the out-of-sample data).

    \begin{figure}[!t]
        \centering
        \includegraphics[width=1.0\textwidth]{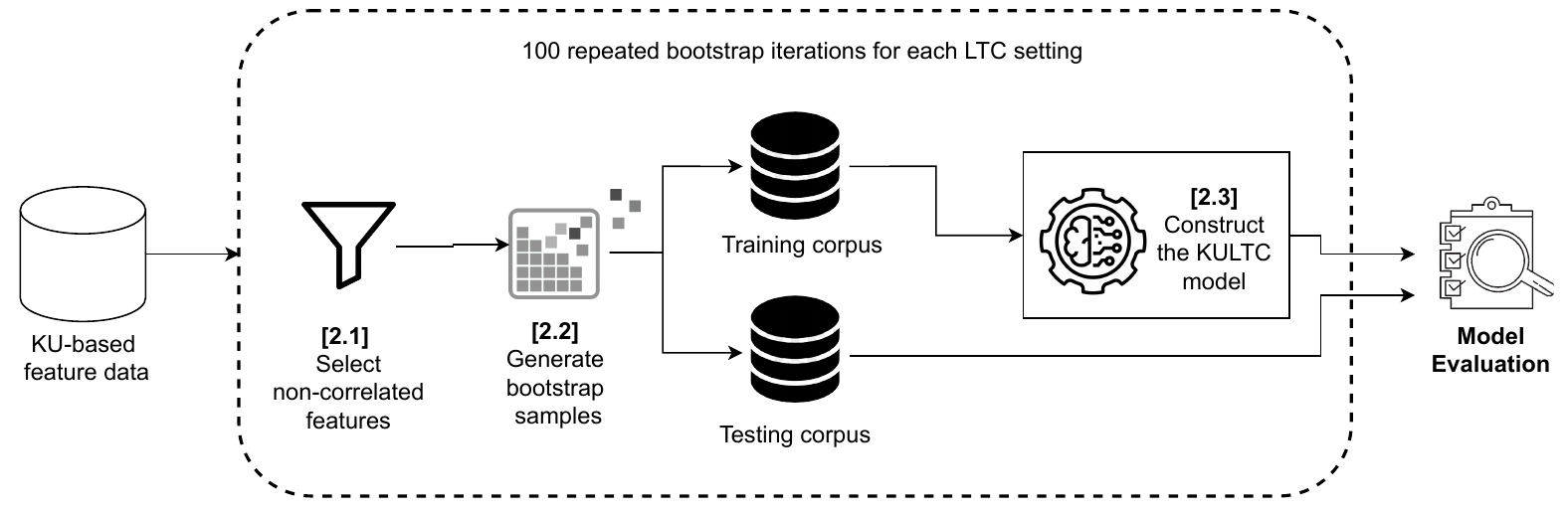}
        \caption{An overview of building the KULTC model.} 
        \label{fig:model-construction}
    \end{figure}

    \item \textit{Step 3) Construct the KULTC model.} 
    We use the non-correlated KU-based features to construct our KULTC model. Since we wish to compare KULTC to the model designed by \citet{bao2019large} (BAOLTC), we follow their machine learning algorithm choice: a random forest classifier with default parameters values (scikit-learn python implementation). This parameter setting of random forest classifier is identified as the optimal choice for achieving the best performance in predicting LTCs compared to the other classifiers that they studied.

\end{itemize}

\subsection{KULTC Model Evaluation}
\label{model_evaluation}
    
To evaluate the accuracy of the KULTC prediction model, we calculate the \textit{Area Under the receiver operator characteristic Curve} (AUC) as it is a robust and threshold-independent measure \citep{Chak_ICSE_SEIP}. AUC score ranges from 0 to 1, where a score of 0.5 indicates that the performance is no better than random guessing. The AUC scores closer to one indicate a model's strong ability to distinguish the class of interest (LTC) from the other class (Non-LTC). Prior studies consider an AUC value of 0.70 to be a decent performance score for a prediction model~\citep{lessmann2008benchmarking,nam2015clami,jiarpakdee2019impact}. Since we are employing an \textit{out-of-sample bootstrap} model validation with 100 repetitions, we obtain 100 AUC values for each prediction model that we evaluate. 

\begin{footnotesize} 
      \begin{mybox}{Summary}
        \begin{itemize}[itemsep = 3pt, label=\textbullet, topsep = 0pt, wide = 0pt]
            \item Data source: GitHub and GHTorrent
            \item Commit data collection date: September 3rd, 2023
            \item GHTorrent snapshot date: March 06, 2021
            \item Number of studied Java projects: 75
            \item Number of studied developers: 1,410
            \item Pieces of collected data: java source code, commit information, and pull requests
            \item Our LTC prediction model (KULTC): A random forest classifier built with KU features that capture developer's expertise in the studied and previous projects, the developer's collaborators expertise in the studied projects, and characteristics of the studied and previous projects.
        \end{itemize}
    \end{mybox}
\end{footnotesize}

    \section{A study of predicting long time contributors using KUs}
\label{sec:Main_study}

In the following, we address our four research questions. For each research question, we discuss our motivation for studying it, the approach used to answer it, and the findings that we observe.

\subsection{RQ1: \RQOne}
\label{sec:KU_Recommendation_System}


\subsubsection{Motivation} 

Previous research highlights that developers primarily join OSS projects to learn and improve their skills~\citep{gerosa2021shifting,lakhani2003hackers}. Our KUs are effective at capturing developers' skills and expertise as well as the characteristics of projects, specifically in terms of programming languages. This makes KUs promising indicators for identifying long-term contributors (LTCs). Therefore, we wish to build a KU-based prediction model (KULTC) for predicting long time contributors. Our objective with this initiative is to evaluate how the KULTC model performs in relation to the baseline for predicting LTCs, aiming to assess its effectiveness.


\subsubsection{Approach}
In Section~\ref{sec:methodology}, we explain our process for constructing our KULTC model. To assess the performance of KULTC, it is essential to establish a baseline LTC prediction model for comparison. In this section, we outline our approach for constructing this baseline model. Additionally, we compare the performance of the KULTC and BAOLTC models and analyze the importance of different feature dimensions.

\smallskip \noindent\textit{\textbf{Construct a baseline model.}} To evaluate the performance of KULTC, we need to select a baseline model for comparison. We find two candidate models in the literature, which we refer to as BAOLTC~\citep{bao2019large} and ZHOULTC~\citep{zhou2014will}. We choose BAOLTC for two reasons: (1) BAOLTC is the most recent between the two, and (2) BAOLTC demonstrates superior performance compared to ZHOULTC. BAOLTC model has 63 different features along five dimensions. The five dimensions are: developer profile, repository profile, developer month activity, repository monthly activity, and collaboration network. These features are detailed in Table~\ref{tab:bao_model_feature}. We carefully extract these features from GHTorrent dataset for the studied developers and projects. Following the steps outlined in Section~\ref{kultc_construction}, we construct the BAOLTC model for each of the studied LTC settings. 

\smallskip \noindent\textit{\textbf{Compare models' performance.}} To determine if there is a significant performance gap between two models (KULTC and BAOLTC), we perform a statistical test. We use a two-sided Wilcoxon signed-rank test, which is a non-parametric statistical test that is suitable for paired data. We consider that there is a statistically significant difference between the distributions when the p-value yielded by the test is less than or equal to 0.05 (i.e., $alpha = 0.05$). Additionally, we employ Cliff's delta ($d$) as an effect size measure to gauge the practical difference between the distributions. We use the following thresholds for interpreting $d$: \textit{negligible} if $|d| \le 0.147$, \textit{small} if $0.147 < |d| \leq 0.33$, \textit{medium} if $0.33 < |d| \leq 0.474$, and \textit{large} otherwise \citep{Romano06}. 

To assess the enhancement in a model's performance compare to the performance of a baseline, we calculate the normalized AUC improvement metric. This metric quantifies the degree of improvement made by the model relative to the maximum possible improvement. We calculate the normalized AUC improvement metric using the following equation:
\vspace{-0.1cm}
\begin {align*}
        & \scriptstyle Normalized\ AUC\ improvement = \frac{AUC\ of\ Proposed Model - AUC\ of\ Baseline Model}{1 - AUC\ of\ Baseline Model} \times 100 \%  
\end{align*}
Here, 1 represents the highest possible score for the AUC. Let's say, we want to calculate the normalized AUC improvement of KULTC over BAOLTC. The AUC of KULTC is 0.81 and the AUC of BAOLTC is 0.75. Therefore, the normalized AUC improvement is (0.81-0.75) / (1 - 0.75) $\times$ 100 \% = 24\%, indicating that KULTC's performance is 24\% closer to the maximum possible improvement compared to the baseline. Since we construct 100 bootstrap models, we calculate the average of the normalized AUC improvement of these models to demonstrate the overall performance improvement.

\subsubsection{Findings}

\begin{figure}[!t]
    \centering
    \includegraphics[width=1.0\textwidth]{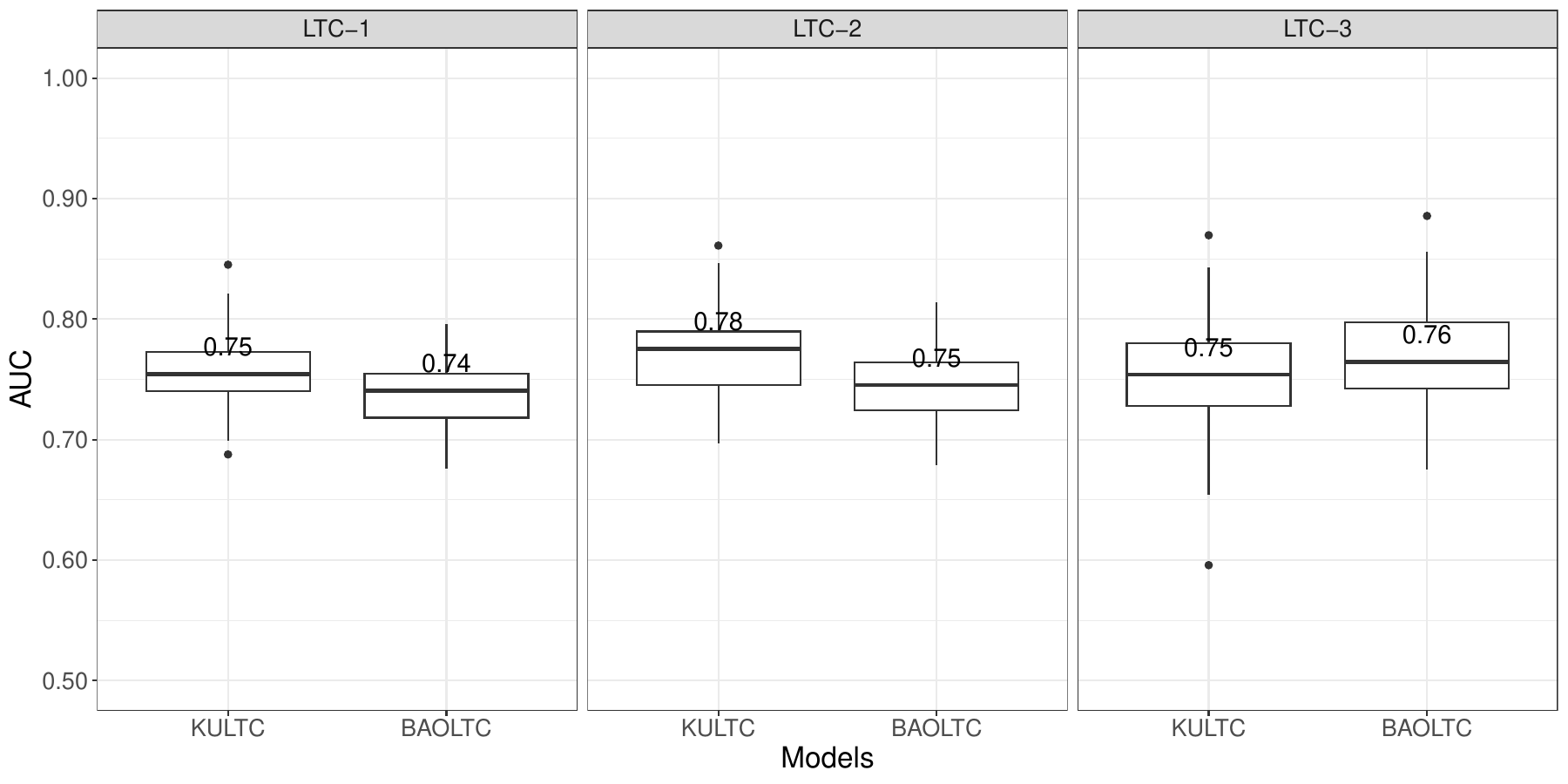}
    \caption{The distribution of AUC of KULTC and BAOLTC across LTC settings.} 
    \label{fig:model_auc}
\end{figure}

\smallskip \observation{The KULTC model outperforms BAOLTC in LTC-1 and LTC-2 settings, while performing similarly in LTC-3.} Figure~\ref{fig:model_auc} presents the distribution of AUC for KULTC and BAOLTC models across three settings for predicting LTCs. We observe that the median AUC of KULTC model is at least 0.75 in all three settings for LTCs. An AUC value of a prediction model equal or higher than 0.70 is considered to be decent and ready for use in practice~\citep{lessmann2008benchmarking,nam2015clami,jiarpakdee2019impact}. The difference between the AUC values of KULTC and BAOLTC is statistically significant (p-value $\leq$ 0.05) for LTC-1 and LTC-2 with a medium effect size (the Cliff's delta value is 0.38 for LTC-1 and 0.44 for LTC-2). Thus, KULTC outperforms BAOLTC for LTC-1 and LTC-2. The normalized AUC improvement of KULTC model over BAOLTC is 6.8\% for LTC-1 and 9.1\% for LTC-2. For LTC-3, we observe that the effect size is negligible (the Cliff's delta value is 0.13) indicating that both models (KULTC and BAOLTC) perform similarly. 

\begin{footnotesize}
    \begin{mybox}{Summary}
    	\textbf{RQ1: \RQOne}
        \tcblower
    	Our KULTC model outperforms the state-of-the-art prediction model (BAOLTC) even though it focuses exclusively on the programming language perspective. In particular:
    	\begin{itemize}[itemsep = 3pt, label=\textbullet, wide = 0pt]
            \item The median AUC of KULTC model is at least 0.75 across all time intervals for LTCs.
            
            \item KULTC model significantly outperforms BAOLTC with a medium effect size for LTC-1 and LTC-2. The normalized AUC improvement of KULTC model over BAOLTC is 6.8\% for LTC-1 and 9.1\% for LTC-2.
            
            \item For LTC-3, both models perform similarly.
    	\end{itemize}
    \end{mybox}
\end{footnotesize}

\subsection{RQ2: \RQTwo}
\label{sec:section_rq2}

\subsubsection{Motivation}

In our study, we utilize five distinct dimensions of KU-based features to construct our KULTC model. Understanding which dimension most significantly impacts the model's ability to predict long time contributors is important, as it allows us to better understand the phenomenon and its drivers. From an operational perspective, identifying key feature dimensions not only streamline the feature engineering process but also enhance the model's practicality for deployment in real-world scenarios.

\subsubsection{Approach} 
\begin{figure}[!h]
    \centering
    \includegraphics[width=1.0\textwidth]{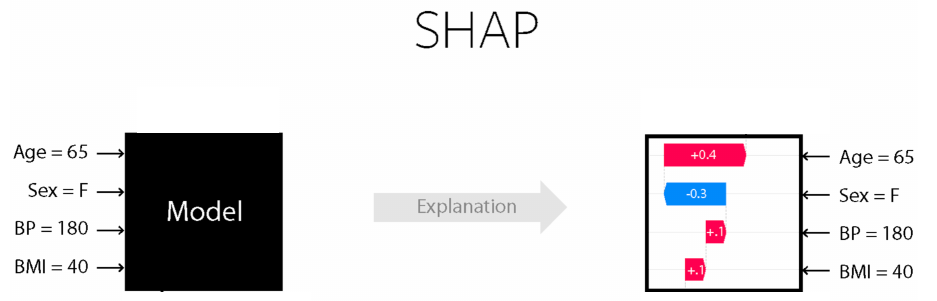}
    \caption{An example illustrating SHAP values for a prediction model.}
    \label{fig:shap_explain}
\end{figure}

To determine the feature dimension that most significantly influence predictions, we conduct a model interpretation analysis utilizing SHAP (SHapley Additive exPlanation). SHAP is a widely used method for model interpretation that builds on game theory to provide an optimal framework for understanding the contribution of each feature to a model's predictions~\citep{molnar2020interpretable}.  We use the \texttt{SHAP} package of \texttt{python} to perform the SHAP analysis \citep{shapper_package}. In practice, the absolute SHAP value for each feature indicates the extent to which that particular feature influences (shifts) the prediction. The higher the absolute SHAP value of a feature, the higher the importance of that feature. To better illustrate we refer Figure~\ref{fig:shap_explain}. A model is built using four features (e.g., Age, Sex, BP and BMI) for predicting heart disease. The colored bars on the right side of Figure~\ref{fig:shap_explain} presents the SHAP values for each feature in the model. The SHAP value of +0.4 for Age indicates a positive influence on prediction, suggesting an increased likelihood of heart disease by 0.4 units. In contrast, the SHAP value of -0.3 for Sex suggests that this feature contributes negatively to the prediction by 0.3 units. Age is considered more influential than Sex because it has a higher absolute SHAP value (|+0.4|>|-0.3|). We use the \texttt{SHAP} package of \texttt{python} to perform the SHAP analysis \citep{shapper_package}. 

We compute feature dimension importance for KULTC as follows. First, we apply SHAP to the KULTC model and calculate the absolute SHAP value for each feature across every record (developer) in our dataset. Next, for each record, we group features by their respective feature dimension and sum their SHAP values. As a result of processing all records, we obtain a distribution of the \textit{sum of SHAP values} for each feature dimension. Finally, we apply the Scott-Knott Effect Size Difference (ESD)~\citep{ghotra_feature_importance_ICSME, mittas_ranking_feature_TSE, kla_model_validation} method to compare (rank) those SHAP distributions. The SK-ESD is a method of multiple comparison that leverages a hierarchical clustering to partition the set of treatment values (e.g., medians or means) into statistically distinct groups with non-negligible difference~\citep{mittas_ranking_feature_TSE}. As a result, we obtain a global feature dimension ranking for the model. We use Tim Menzies's implementation of the Scott-Knott ESD technique, which employs non-parametric statistical tests and the Cliff's Delta measure of effect size \citep{tim_sk}.

\subsubsection{Findings}

\begin{figure}[!t]
    \centering
    \includegraphics[width=1.0\textwidth]{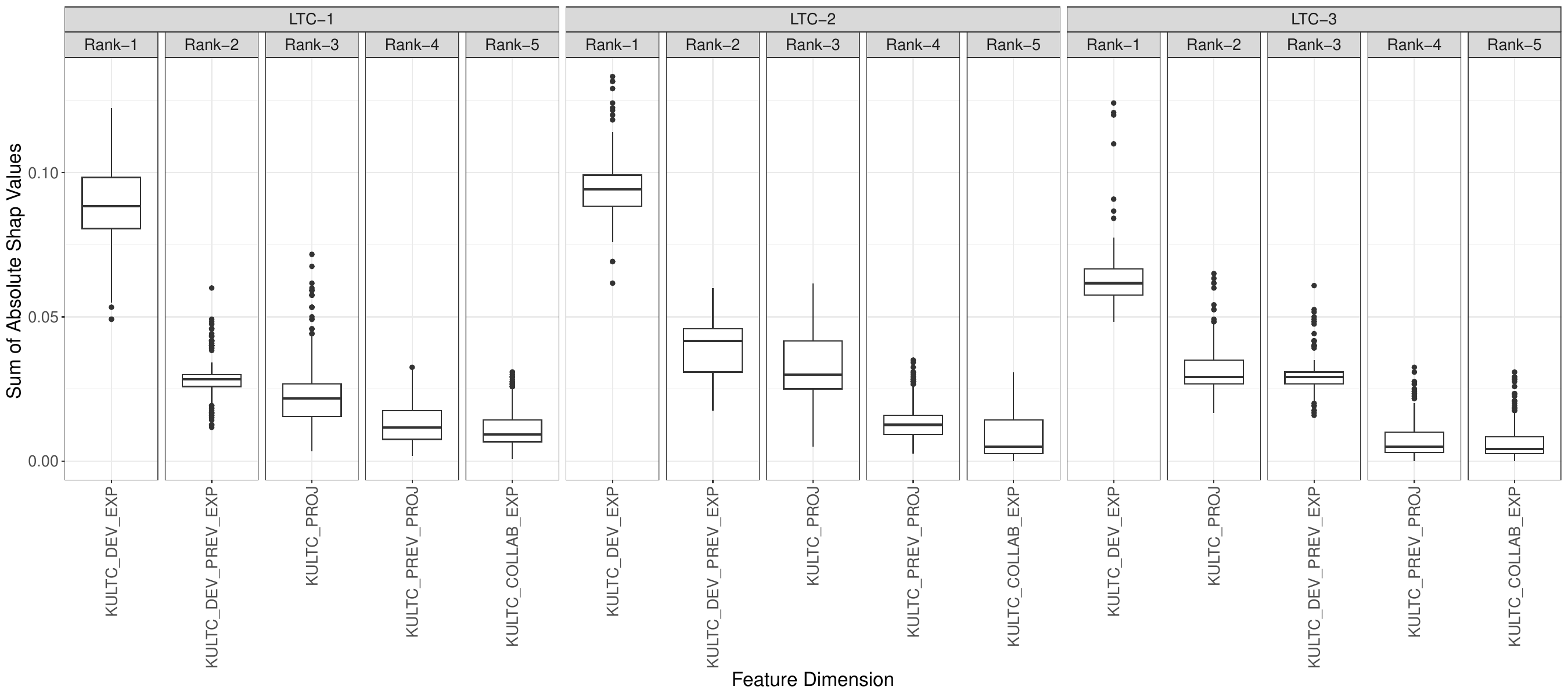}
    \caption{The distribution of the sum of absolute SHAP values for each KU-based feature dimension and their importance ranking in predicting LTCs (lower rank = better).}
    \label{fig:model_feature_analysis}
\end{figure}

\sloppy \smallskip \observation{The KU-based developer expertise feature dimension is the most influential dimension across all settings for predicting LTCs.} Figure~\ref{fig:model_feature_analysis} presents the sum of absolute SHAP values for each KU-based feature dimension. Distributions are grouped according to their rank. We observe that the KU-based feature dimension representing a developer's expertise (KULTC\_DEV\_EXP) during his first month's development activity in the studied project is the top-ranked influential feature dimension for predicting LTCs across all three settings (i.e., LTC-1, LTC-2, and LTC-3). The median for KULTC\_DEV\_EXP is 0.09, which indicates that the group of features within KULTC\_DEV\_EXP impacts the model classification output with 0.09$\times$100\% (9\%) percentage points.

\sloppy Regarding the other two expertise related feature dimensions, developer prior expertise dimension (KULTC\_DEV\_PREV\_EXP) consistently ranks above the collaborator expertise dimension (KULTC\_COLLAB\_EXP). In fact, KULTC\_DEV\_PREV\_EXP is the second-highest ranked feature dimension for LTC-1 and LTC-2, while KULTC\_COLLAB\_EXP is always ranked the lowest.

\sloppy When comparing the impact of KU-based project characteristics dimension on predicting LTCs, we observe that the dimension related to the characteristics of the studied projects (KULTC\_PROJ) has more impact than the dimension related to the characteristics of previous projects (KULTC\_PREV\_PROJ).

\begin{footnotesize}
    \begin{mybox}{Summary}
    	\textbf{RQ2: \RQTwo}
        \tcblower
        The KU-based feature dimension representing a developer's expertise during his first month's development activity (KULTC\_DEV\_EXP) is the most influential feature dimension of KULTC for predicting LTCs.
    \end{mybox}
\end{footnotesize}

\subsection{RQ3: \RQThree}
\label{sec:section_rq3}

\subsubsection{Motivation} 

This research question aims to understand whether the performance of the KULTC model for predicting LTCs can be enhanced by integrating it with features from an existing model. The motivation stems from the hypothesis that blending diverse sets of features might capture a broader spectrum of factors that affect prediction of LTCs, thereby possibly resulting in more accurate forecasts of who will become an LTC.

\subsubsection{Approach} 

\sloppy We build a combined model named KULTC+BAOLTC for predicting LTCs using all studied features of KULTC and all features of BAOLTC. Following the outlined steps in Section~\ref{kultc_construction}, we build the KULTC+BAOLTC model for all studied settings for predicting LTCs. We compare the performance of the combined KULTC+BAOLTC model with the performance of KULTC and BAOLTC. We rank the studied prediction models (KULTC, BAOLTC and KULTC+BAOLTC) using SK-ESD~\citep{ghotra_feature_importance_ICSME, mittas_ranking_feature_TSE, kla_model_validation}. We also analyze the feature dimension importance of the KULTC+BAOLTC model using the same SHAP analysis method explained in Section~\ref{sec:section_rq2}.

\subsubsection{Findings}

\begin{figure}[!t]
    \centering
    \includegraphics[width=1.0\textwidth]{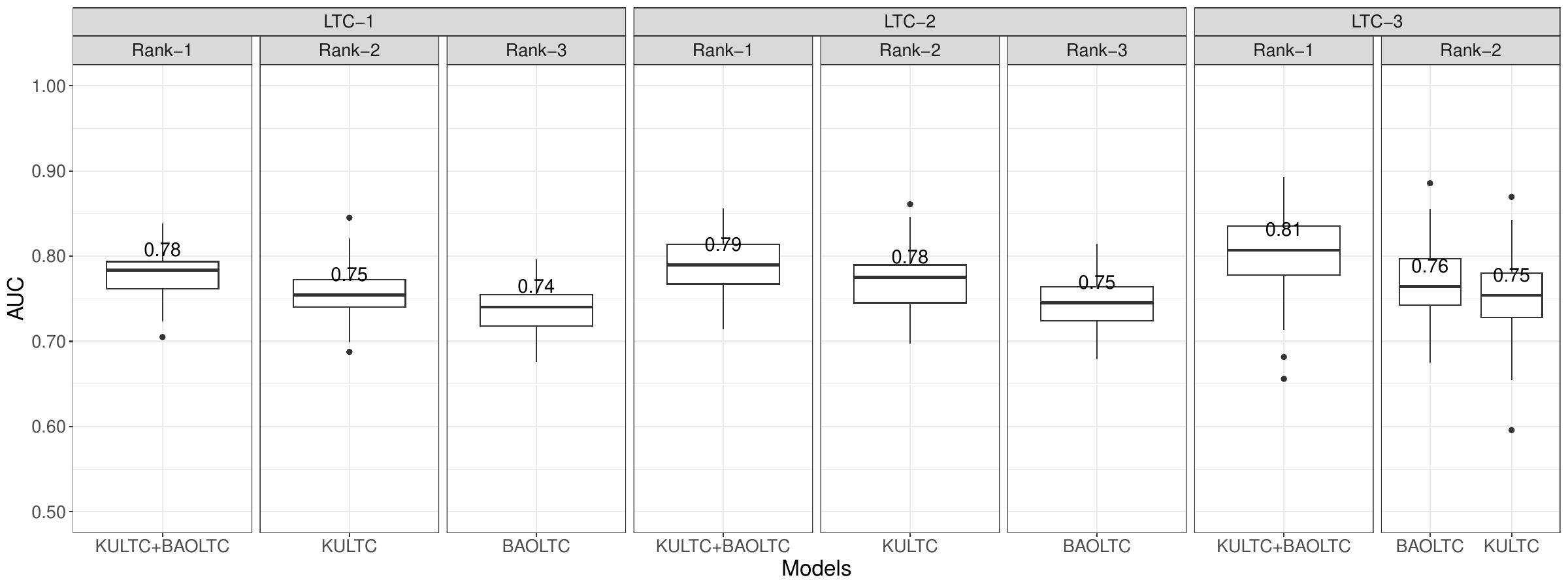}
    \caption{The distribution of AUC of KULTC, BAOLTC and the combined model KULTC+BAOLTC. Models are grouped according to their Scott-Knott ESD rank (lower rank = better).} 
    \label{fig:combined_model_result}
\end{figure}

\sloppy \observation{Combining KU-based features with BAOLTC's features results in a higher performing model for predicting LTCs.} Figure~\ref{fig:combined_model_result} depicts the performance of the three studied models. We observe that KULTC+BAOLTC is the top performing model, surpassing both the individual BAOLTC and KULTC across all settings for predicting LTCs. Specifically, in the LTC-3 setting, KULTC+BAOLTC achieves the highest median AUC (0.81), representing a 14.5\% improvement over BAOLTC and a 19.2\% improvement over KULTC. Additionally, for LTC-1 and LTC-2, the performance improvements of KULTC+BAOLTC over BAOLTC are 15.6\% and 16.5\%, respectively. We thus conclude that KULTC+BAOLTC is a more effective model for predicting LTCs than either of the individual models due to its more complete feature set.

\begin{figure}[!t]
    \centering
    \includegraphics[width=1.0\textwidth]{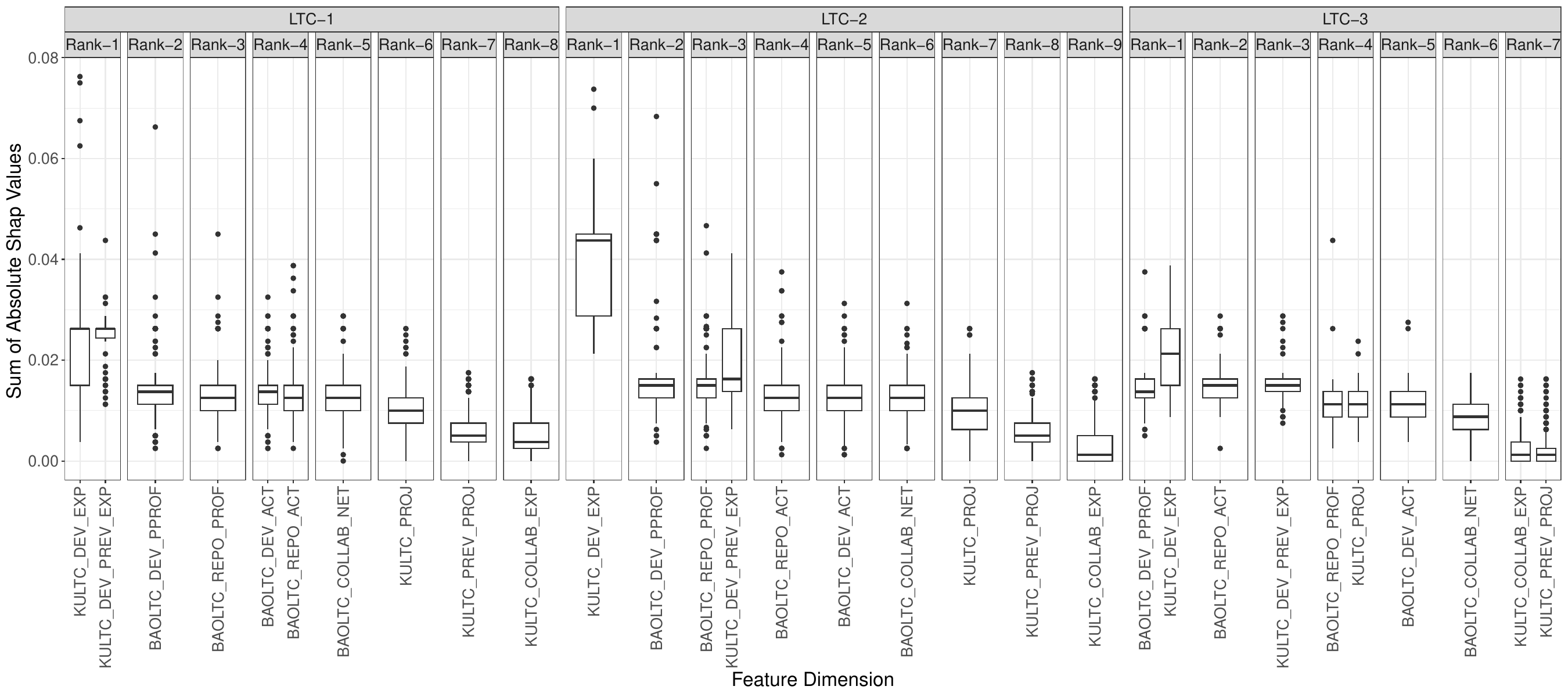}
    \caption{The distribution of the sum of absolute SHAP values for each feature dimension and their importance ranking in predicting LTCs (lower rank = better).} 
    \label{fig:combined_model_feature_importance}
\end{figure}

\sloppy \observation{The KU-based developer expertise in the studied projects stands out as the most influential feature dimension for predicting LTCs in the KULTC+BAOLTC model.} Figure~\ref{fig:combined_model_feature_importance} presents the distribution of the sum of absolute SHAP values for each feature dimension of the KULTC+BAOLTC model. We observe that KULTC\_DEV\_EXP, the KU-based feature dimension representing a developer's programming language expertise during his first month's development activity in the studied projects, is the top ranked one across all settings for predicting LTCs. In other words, the programming language expertise that developers demonstrate within one month (a short period of time) has the strongest influence in predicting LTCs. Interestingly, the BAOLTC\_DEV\_ACT feature dimension, which captures the first month's development activity of developers (instead of capturing programming language expertise), ranks only forth (LTC-1) or fifth (LTC-2 and LTC-3). 

\sloppy We also make the following observations. The second most influential KU dimension is KULTC\_DEV\_PREV\_EXP, which refers to developers programming language experience in prior projects. The remaining KU feature dimensions rank much higher, but still influence the model's classification (note the non-zero median). Among the feature dimensions from the original BAOLTC model (BAOLTC\_*), the developer profile dimension (BAOLTC\_DEV\_PROF) is the most influential one.


\begin{footnotesize}
    \begin{mybox}{Summary} 
    	\textbf{RQ3: \RQThree}
        \tcblower
        Yes. We combined the features from KULTC and BAOLTC into a new combined model that we call KULTC+BAOLTC. Such a combined model outperforms the original models with a non-negligible effect size. In particular,
    	\begin{itemize}[itemsep = 3pt, label=\textbullet, wide = 0pt]
            \item KULTC+BAOLTC ranks first across all settings for predicting LTCs (i.e., LTC-1, LTC-2, and LTC-3) with a median AUC that ranges from 0.78 to 0.81.
            \item The normalized AUC improvement of KULTC+BAOLTC over BAOLTC is 15.6\% for LTC-1, 16.5\% for LTC-2 and 14.5\% for LTC-3. 
            \item In the KULTC+BAOLTC model, the KU-based feature dimension that represents developer expertise in the studied projects is the most influential one for predicting LTCs.
    	\end{itemize} 
    \end{mybox}
\end{footnotesize}
\subsection{RQ4: \RQFour}
\label{sec:section_rq4}

\subsubsection{Motivation} 

We use a random forest classifier with default hyper-parameter values as the standard configuration for building the combined model KULTC+BAOLTC (Section~\ref{sec:section_rq3}). However, it is important to note that the performance of a model can vary significantly across different classifiers, as each has its own strengths and weaknesses that suit different types of data and problems~\citep{Chak_ICSE_SEIP}. Furthermore, optimal performance is often achieved by tuning specific hyperparameters of a classifier~\citep{tantithamthavorn2016automated, liao2022empirical, wong2019can}. Adjusting these parameters can significantly enhance a model's ability to learn from the training data both effectively and efficiently. Hence, in this RQ we determine whether a more accurate version of KULTC+BAOLTC can be obtained by simply replacing our random forest with an alternative hyper-parameter-tuned classifier.




\subsubsection{Approach} 

To assess the impact of different classifiers and hyper-parameter tuning on KULTC+BAOLTC, we select a diverse set of classifiers: K-Nearest Neighbor (KNN), Naive Bayes (NB), Decision Trees (DT), Random Forest (RF), XGBoost (XGB), and LightGBM (LGBM). The latter two, XGB and LGBM, represent the most recent and advanced classifiers, while the others are well-established and traditionally popular options in empirical software engineering studies. 

To generate bootstrap samples, we apply the out-of-sample bootstrap model validation technique with 100 repetition. For each iteration, we have a bootstrap sample that we use to train the model. We test the model with the data that do not appear in the sample (i.e., the out-of-sample data).

\sloppy To determine the most effective model configuration, we utilize Scikit-learn’s GridSearchCV\footnote{\url{https://scikit-learn.org/stable/modules/generated/sklearn.model_selection.GridSearchCV.html}}. The  GridSearchCV exhaustively searches through a specified grid of hyper-parameters (param\_grid) and employs cross-validation. The hyper-parameter configurations for the selected classifiers that we explore in our study are detailed in Table~\ref{tab:hyper_parameter_config}. We apply the  GridSearchCV on the bootstrap sample data using a 10-fold cross validation. Then, the GridSearchCV identifies and returns the model with the best performance based on the hyper-parameter tuning.  We then test this model using the out-of-sample data. To evaluate the model, we apply the same approach that we describe in Section~\ref{model_evaluation}.

\begin{table}[!ht]
    \caption{The studied classification algorithms with different parameter settings for hyper parameter optimization.}
    \label {tab:hyper_parameter_config}
    \resizebox{\columnwidth}{!}{
    \begin{tabular}{@{}lllll@{}}
    \toprule
    \textbf{\begin{tabular}[c]{@{}l@{}}Classifier\\ Name\end{tabular}} & \textbf{\begin{tabular}[c]{@{}l@{}}Classification\\ Algorithm\end{tabular}} & \textbf{\begin{tabular}[c]{@{}l@{}}Parameter\\ Name\end{tabular}} & \textbf{\begin{tabular}[c]{@{}l@{}}Parameter\\ Description\end{tabular}}                                                                                                                  & \textbf{\begin{tabular}[c]{@{}l@{}}Studied candidate\\ parameter values\end{tabular}} \\ \midrule
    KNN                                                                & \begin{tabular}[c]{@{}l@{}}K-Nearest \\ Neighbour\end{tabular}              & n\_neighbors                                                      & \begin{tabular}[c]{@{}l@{}}The number of neighbors  \\ required for  each sample\end{tabular}                                                                                             & \{1, 5, 9, 13, 17, 20\}                                                               \\ \midrule
    NB                                                                 & Naive Bayes                                                                 & var\_smoothing                                                    & \begin{tabular}[c]{@{}l@{}}The portion of the largest variance \\ of all features  that is added to \\ variances  for calculation stability.\end{tabular}                                 & \{1e-5, 1e-9, 1e-11, 1e-15\}                                                          \\ \midrule
    \multirow{3}{*}{DT}                                                & \multirow{3}{*}{Decision Tree}                                              & criterion                                                         & \begin{tabular}[c]{@{}l@{}}The function to measure the quality \\ of a split.\end{tabular}                                                                                                & \{`gini', `entropy', `log\_loss'\}                                                    \\ \cmidrule{3-5}
                                                                       &                                                                             & max\_depth                                                        & The maximum depth of the tree.                                                                                                                                                            & \{None, 5, 10\}                                                                       \\ \cmidrule{3-5}
                                                                       &                                                                             & ccp\_alpha                                                        & \begin{tabular}[c]{@{}l@{}}Complexity parameter used for \\ Minimal  Cost-Complexity Pruning.\end{tabular}                                                                                & \{0.0001, 0.001, 0.01, 0.1, 0.5\}                                                     \\ \midrule
    \multirow{2}{*}{RF}                                                & \multirow{2}{*}{Random Forest}                                              & n\_estimator                                                      & The number of trees in the forest.                                                                                                                                                        & \{10, 50, 100, 200\}                                                                  \\ \cmidrule{3-5}
                                                                       &                                                                             & max\_depth                                                        & \begin{tabular}[c]{@{}l@{}}The maximum depth of the tree. \\ If None,  then  nodes  are expanded \\ until all leaves  are pure.\end{tabular}                                              & \{None, 5, 10\}                                                                       \\ \midrule
    \multirow{3}{*}{XGB}                                               & \multirow{3}{*}{XGBoost}                                                    & n\_estimator                                                      & \begin{tabular}[c]{@{}l@{}}The number of boosting rounds or \\ trees  to build.\end{tabular}                                                                                              & \{10, 50, 100, 200\}                                                                  \\ \cmidrule{3-5}
                                                                       &                                                                             & max\_depth                                                        & The maximum depth of the tree.                                                                                                                                                            & \{None, 5, 10\}                                                                       \\ \cmidrule{3-5}
                                                                       &                                                                             & learning\_rate                                                    & \begin{tabular}[c]{@{}l@{}}This parameter controls how much \\ the  model is  adjusted  in response\\ to the  estimated error each  time \\ the model   weights are updated.\end{tabular} & \{0.1, 0.01, 0.001\}                                                                  \\ \midrule
    \multirow{3}{*}{LGBM}                                              & \multirow{3}{*}{LightGBM}                                                   & n\_estimator                                                      & The number of boosted trees to fit.                                                                                                                                                       & \{10, 50, 100, 200\}                                                                  \\ \cmidrule{3-5}
                                                                       &                                                                             & num\_leaves                                                       & \begin{tabular}[c]{@{}l@{}}The maximum tree leaves for \\ base learners.\end{tabular}                                                                                                     & \{None, 5, 10\}                                                                       \\ \cmidrule{3-5}
                                                                       &                                                                             & learning\_rate                                                    & \begin{tabular}[c]{@{}l@{}}The Boosting learning rate that \\ controls  how quickly the model \\ adjusts to the  error  in each  \\ iteration of training\end{tabular}                    & \{0.1, 0.01, 0.001\}                                                                  \\ \bottomrule
    \end{tabular}
    }
    \end{table}

We compare the performance of different KULTC+BAOLTC models that are built using the selected classifiers, by following the same ranking method that we describe in Section~\ref{sec:section_rq3}.

\subsubsection{Findings}

\begin{figure}[!t]
    \centering
    \includegraphics[width=1.0\textwidth]{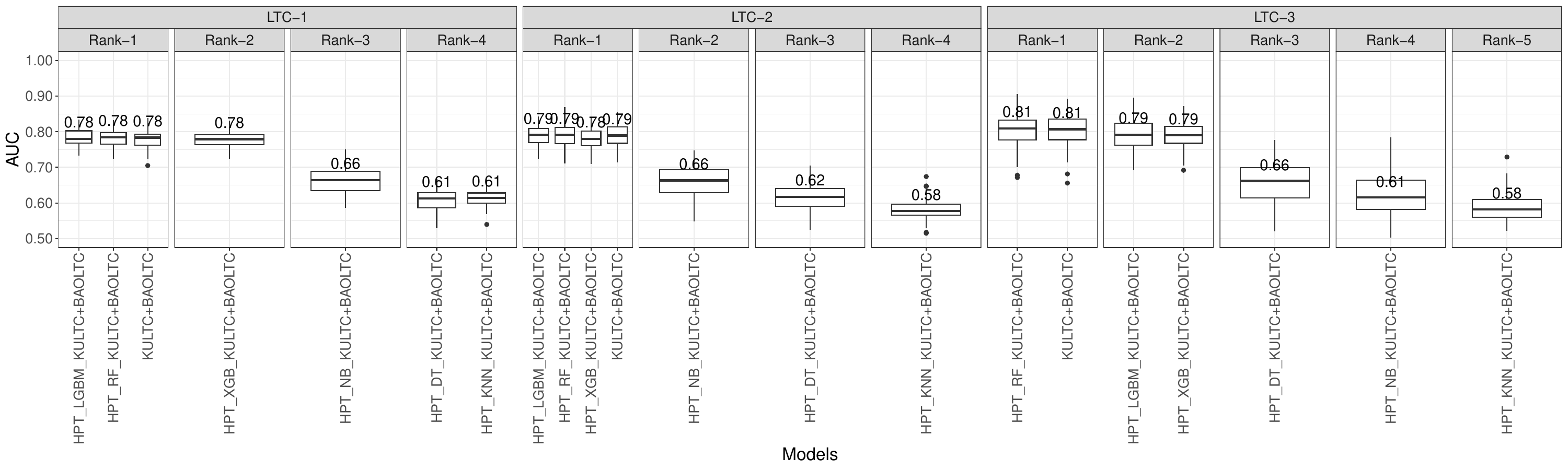}
    \caption{The AUC distribution hyper-parameter-tuned models and our original KULTC+BAO (random forest with default hyper-parameter values). The names of the models that have gone hyper-parameter tuning begin with ``HPT\_X,'' where \textit{X} represents the name of one of the classifiers we studied (lower rank = better).} 
    \label{fig:combined_model_hpt}
\end{figure}

\observation{Our original KULTC+BAOLTC remains the top-performing model across all settings for predicting LTCs, outperforming all other hyper-parameter tuned classifiers.}  Figure~\ref{fig:combined_model_hpt} presents the AUC distribution of KULTC+BAOLTC and the hyper-parameter-tuned variations. We observe that KULTC+BAOLTC and the hyper-parameter-tuned random forest model (HPT\_RF\_KULTC+BAOLTC) perform similarly and lead the ranking across all settings for predicting LTCs. Models built using traditional classifiers (e.g., NB, DT and KNN) achieved suboptimal performance with a median AUC below 0.70. Among those, the KNN-based model performs the worst, with a median AUC below 0.60. In contrast, more advanced classifiers such as XGBoost and LightGBM achieve better results, with median AUC scores above 0.75. Yet, those do not surpass the performance of KULTC+BAOLTC. Hence, we cannot improve KULTC+BAOLTC by simply switching to a different classifier with tuned hyper-parameters.


\begin{footnotesize}
    \begin{mybox}{Summary}
    	\textbf{RQ4: \RQFour}
        \tcblower
        We cannot improve the accuracy of KULTC+BAOLTC by simply using a different classifier with tuned hyper-parameters. 
    \end{mybox}
\end{footnotesize}
\subsection{RQ5: \RQFive}

\subsubsection{Motivation} 

In the development of predictive models, using a wide array of features often leads to high performance. However, the process of extracting a full spectrum of features can be challenging and expensive, involving extensive data collection and processing efforts. For example, engineering features related to developer prior expertise needs a detailed analysis of all previous projects of developers. Given these constraints, there is a compelling need to identify a more cost-effective model that utilizes a minimal number of features while still delivering good performance.

\subsubsection{Approach}
To build a cost-effective model, we focus on a specific feature dimension from the KULTC model: the KU-based developer expertise in the studied  projects (KULTC\_DEV\_EXP). Our rationale is that engineering this KU-based feature dimension does not require an extensive data collection approach. Instead, it only involves analyzing the developer's commits made within the first 30 days from their initial commit to the studied projects. Additionally, it simplifies the process for detecting KUs from the source code, since there are fewer code snapshots to mine and analyze. 

\sloppy We combine KULTC\_DEV\_EXP feature dimension with all features of BAOLTC, which can be easily extracted from the GHTorrent dataset. This integration of feature dimensions forms a new combined model called KULTC\_DEV\_EXP+BAOLTC. To build the KULTC\_DEV\_EXP+BAOLTC model, we follow the approach for model construction presented in Section~\ref{sec:kultc-feateng}. We determine the best classifier (with hyper-parameter tuning) that performs the best in predicting LTCs by applying the same approach described in RQ4. Then, we evaluate KULTC\_DEV\_EXP+BAOLTC and compare the performance of it with that of KULTC, BAOLTC and KULTC+BAOLTC models. The comparison involves ranking the models according to the same SK-ESD method detailed in Section~\ref{sec:section_rq3}. 

 
\subsubsection{Findings} 

\begin{figure}[!ht]
    \centering
    \includegraphics[width=1.0\textwidth]{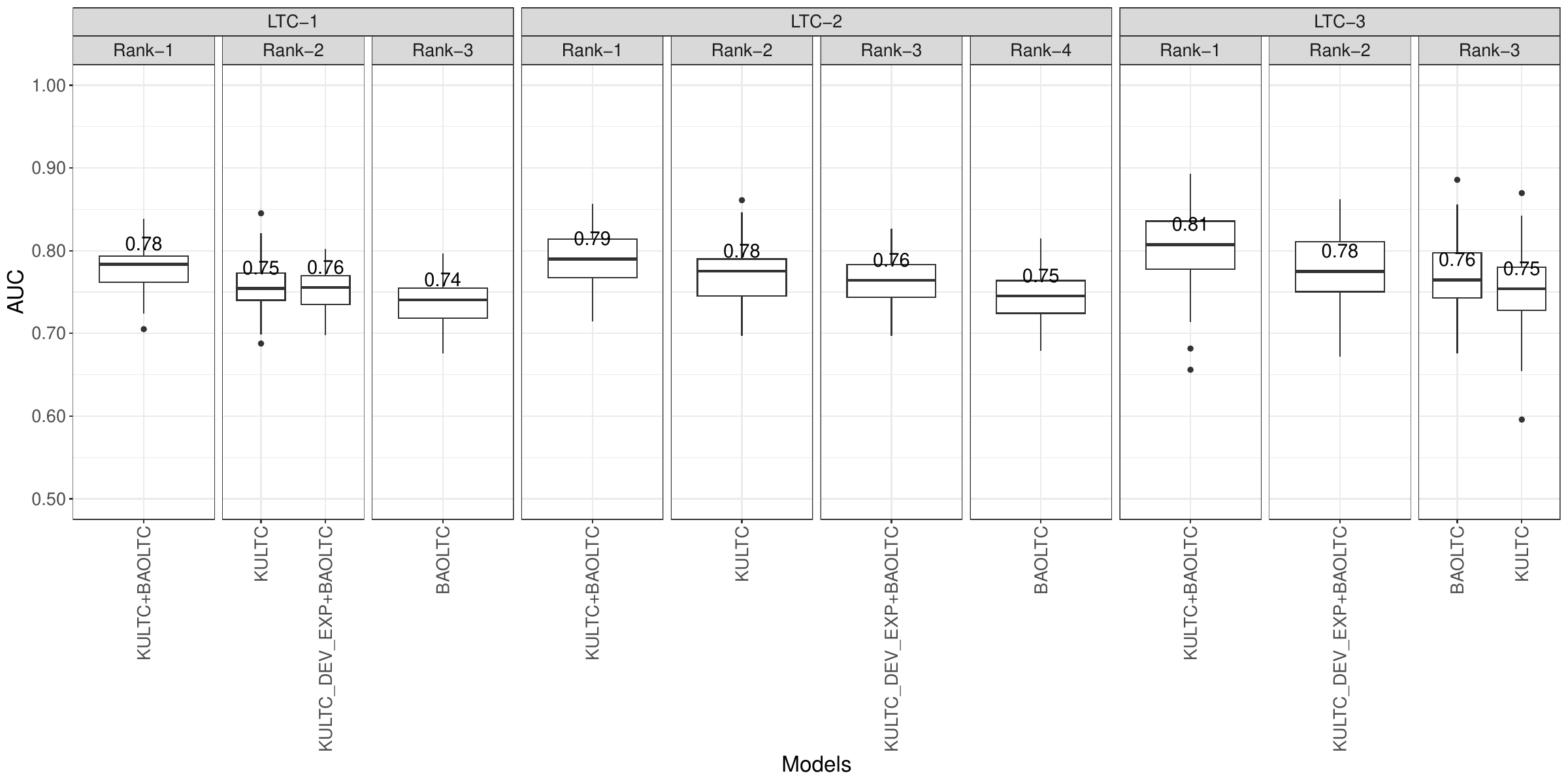}
    \caption{The AUC distribution of the cost-effective KULTC\_DEV\_EXP+BAOLTC, as well as KULTC, BAOLTC and KULTC+BAOLTC. Models are grouped according to their Scott-Knott ESD rank (lower rank = better).} 
    \label{fig:cost_effective_model_comparison}
\end{figure}

\sloppy \observation{The random forest classifier with default parameter settings results in the best cost-effective KULTC\_DEV\_EXP+BAOLTC that outperforms BAOLTC (analogous to KULTC+BAOLTC).} Figure~\ref{fig:cost_effective_model_comparison} presents the AUC distribution of the KULTC\_DEV\_EXP+BAOLTC, KULTC, BAOLTC, and KULTC+BAOLTC models. Upon rerunning the approach from RQ4, we observe that random forest with default parameter settings is the best classifier that we refer to the cost effective model as KULTC\_DEV\_EXP+BAOLTC. We observe that the KULTC\_DEV\_EXP+BAOLTC consistently outperforms BAOLTC across all settings for predicting LTCs. For LTC-3, KULTC\_DEV\_EXP+BAOLTC excels in predicting LTCs, achieving a median AUC of 0.78, which outperforms both KULTC and BAOLTC. The normalized AUC improvement of KULTC\_DEV\_EXP+BAOLTC over BAOLTC is 5.3\% for LTC-1 , 6.1\% for LTC-2 and 3.3\% for LTC-3. We note that when budget constraints are not an issue, KULTC+BAOLTC is the recommended choice as it consistently remains the top-performing model across all LTC settings.

\begin{figure}[!t]
    \centering
    \includegraphics[width=1.0\textwidth]{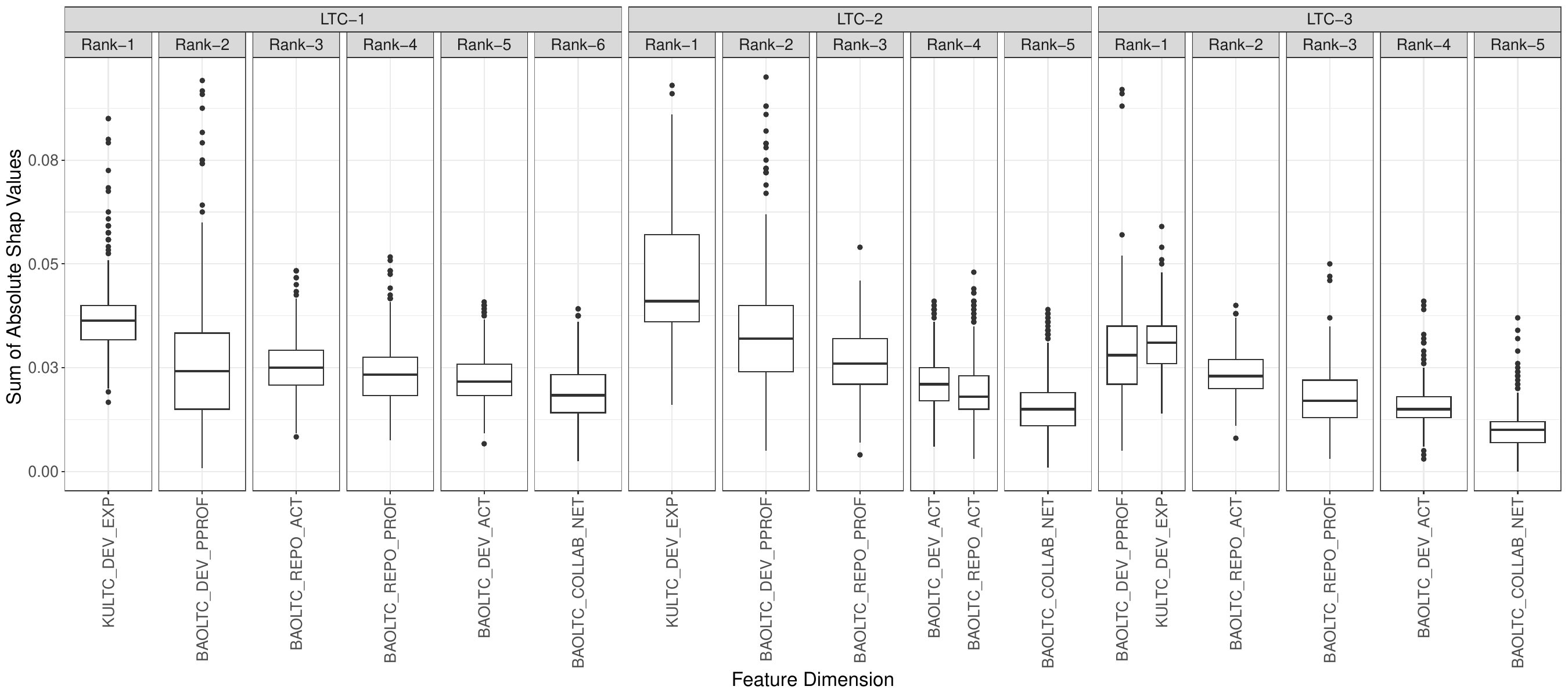}
    \caption{The distribution of the sum of absolute Shapley values for each feature 
    dimension and their importance ranking in predicting LTCs}
    \label{fig:cost_effective_model_feature_importance}
\end{figure}

\observation{The KU-based developer expertise in the studied projects remains the most influential feature dimension in the cost-effective  model.} Figure~\ref{fig:cost_effective_model_feature_importance} depicts the \textit{sum of absolute SHAP values} for each feature dimension of the cost-effective model (KULTC\_DEV\_EXP+BAOLTC). We observe that KULTC\_DEV\_EXP ranks first across all LTC settings. Such a result underscores the significant role of the KU-based feature dimension related to developer expertise in the studied projects for predicting LTCs. Among the BAOLTC feature dimensions, the developer profile dimension (BAOLTC\_DEV\_PROF) is the most influential one, ranking second in LTC-1 and LTC-2 settings, while the collaboration network feature dimension (BAOLTC\_COLLAB\_NET) is the least influential one.

\begin{footnotesize}
    \begin{mybox}{Summary}
    	\textbf{RQ5: \RQFive}
        \tcblower
    	The cost-effective model KULTC\_DEV\_EXP+BAOLTC outperforms the BAOLTC model. In particular,
        \begin{itemize}[itemsep = 3pt, label=\textbullet, wide = 0pt]
            \item The random forest classifier with default parameter settings achieve the best performance for KULTC\_DEV\_EXP+BAOLTC.

            \item The normalized AUC improvement of the KULTC\_DEV\_EXP+BAOLTC over BAOLTC is 5.3\% for LTC-1 , 6.1\% for LTC-2 and 3.3\% for LTC-3.

            \item The only KU-based feature dimension in the cost-effective model (i.e., KU-based developer expertise in the studied projects) is the most influential feature dimension across all LTC settings.
            

    	\end{itemize} 
    \end{mybox}
\end{footnotesize}
	\section{Discussion}
\label{sec:Discussion}

In this section, we discuss two crucial aspects of model construction for predicting LTCs: balancing model complexity with performance and expanding the scope of KUs.

\noindent \textbf{Trade-off between model complexity and performance improvement.} 
We observe that KULTC outperforms BAOLTC in LTC-1 and LTC-2. The performance of the combined model (KULTC+BAOLTC) is even better -- the performance improvement of KULTC+BAOLTC over BAOLTC is more than twice than that of KULTC. However, such performance improvement can cause an overhead in model complexity, specially in the approach of data processing and feature engineering. Collecting features across all dimensions can involve extensive data collection and processing efforts. 
For example, KULTC\_DEV\_PREV\_EXP dimension needs to analyze the code snapshot of all commits that a developer perform in each of the previous projects the developer worked on before contributing to the project under study. This suggests analyzing a large number of commits that can increase operational burdens to engineer this feature. To reduce such operational burdens, we introduce a cost-effective predictive model (KULTC\_DEV\_EXP + BAOLTC) that also outperforms BAOLTC (but the overall improvement is lower than KULTC+BAOLTC). This cost-effective model combines only one feature dimension of KULTC, namely the developer expertise in the studied projects (KULTC\_DEV\_EXP), with the features of BAOLTC. Collecting KULTC\_DEV\_EXP features for a developer is much cheaper since it involves analyzing the code snapshot of only a few commits that developer performs during the first month since their initial commit to the project. Additionally, we include our custom-built KU-detector in the supplementary package, which is ready for use to help researchers and practitioners extract KUs and reduce overhead.

\noindent \textbf{Expanding KUs to include broader forms of expertise.} In this paper, we leverage KUs of programming languages to predict long time contributors. The predictive models KULTC and KULTC+BAOLTC outperform the state-of-the-art BAOLTC model. Our feature dimension importance analysis for these models reveal that developers KU-based programming language expertise in the studied projects is the top-ranked influential feature dimension for predicting LTCs. Such results indicate that KUs are good predictor for LTC prediction models.

Given our encouraging results in predicting LTCs, researchers should explore the applicability of our approach to different programming languages (e.g., python projects) and consider the development of design alternatives, potentially more cost-effective and automated operationalization methods.  One promising avenue could involve utilizing programming language documentation and leveraging large language models (LLMs) to summarize and extract relevant information (e.g., code concepts) efficiently. A natural extension of our work would involve integrating other form of technical knowledge such as libraries and API knowledge. This could include high-level libraries widely used in data science, such as pandas, as well as lower-level, more domain-specific libraries such as the Vulkan API. Incorporating such diverse forms of technical knowledge would likely offer a more comprehensive view of developers' expertise and their potential for sustained engagement in a project. This broader perspective can significantly enhance the predictive power of models aimed at identifying long time contributors.

	\section{Related Work}
\label{sec:Related_Work}

The phenomenon of developer turnover, characterized by the ongoing entry and exit of developers in software development, has been the subject of study for many years. Researchers have explored what motivates developers to participate in and what drives them to leave OSS projects. \citet{yu2012empirical} identified that the key indicators of developer turnover are the objective characteristics of OSS projects and the developers' personal expectations. \citet{schilling2012will} observe that development experience and conversational knowledge significantly correlate with developer retention. \citet{hynninen2010off} carried out a survey among developers and discovered that developer turnover might significantly reflect a lack of commitment. \citet{sharma2012examining} developed a linear logistic model accounting for factors at both the developer and project levels, revealing that past activity, developer role, project size, and project age significantly influence developer turnover. ~\citet{zanatta2017barriers} conducted a study to  uncover obstacles hindering new participants in software-crowdsourcing projects. They identify that insufficient documentation is the primary barrier to their involvement. \citet{yamashita2016magnet} analyzed a vast array of GitHub projects to measure project characteristics from the perspective of developer attraction and retention. They categorized GitHub projects as either `magnet' or `sticky' based on developer movement within the platform. Their findings reveal that 23\% of developers continue contributing to the same project, with larger projects having a higher likelihood of drawing in and retaining developers.

\citet{iaffaldano2019developers} study the break patterns of developers in OSS. The authors investigate temporary and permanent breaks that developers take from contributing. In a further study by \citep{calefato2022will}, the authors investigate OSS developers' lifecycle focusing on their breaks and disengagement. Through interviews, they develop a state model describing OSS (in)activity states and transitions, which they apply to study the inactivity patterns of core developers in 18 GitHub OSS projects. Their analysis shows that core developers often take breaks of various lengths and types, with all developers taking at least one break, and 97\% moving to noncoding activities like code review and project management. The primary personal reason for developers to return to contributing is their professionalism and sense of responsibility towards the project. \citet{miller2019people} explore the reasons behind established contributors disengaging from open-source projects. Their research indicates that contributors leave for various reasons, most commonly due to personal transitions such as job changes or leaving academia.

A few studies focus on predicting long time contributors~\citep{zhou2014will,bao2019large} which are closely related to our work. \citet{zhou2014will} model an individual's chance to become a long time contributor (LTC) using the logistic regression algorithm. The authors build a logistic regression model based on 10 features across three dimensions: (1) degree of engagement (contributors' activities such as the number of commits), (2) macro-climate (the overall project environment such as the total number of contributors) and (3) micro-climate (the environment unique for each contributor such as the size of the peer group). Their evaluation results show that the precision of their model is 38 times higher than for a random predictor (i.e., a predictor that randomly selects a developer to be an LTC). \citet{bao2019large} study long time contributors for three different settings on the time interval: 1, 2 and 3 years. The authors build machine learning models using 63 features based on the GitHub data which belongs to five different dimensions: developer profile, repository profile, developer monthly activity, repository monthly activity and collaboration network. They apply five different classifiers (such as naive Bayes, SVM, decision tree, kNN and random forest) and evaluate their models on the selected 917 projects from GHTorrent dataset. The evaluation results show that the random forest classifier achieves the best performance.

Our work is different from the above work. The above-mentioned example work mainly focus on process metrics (e.g., number of submitted commits, number of pull requests and number of commits) and basic project characteristics (e.g., total number of commits, programming language, and total number of stars). In contrast, we leverage knowledge units (KUs) of programming language and tailor them to engineer features across five different dimensions. These KU-based feature dimensions capture developers's expertise and project characteristics in terms of how they use programming languages and they are used to build model (KULTC) for predicting LTCs.

    \section{Threats to Validity}
\label{sec:Limitations_And_Threats}

\smallskip \noindent\textbf{Construct validity.} One of the threats to construct validity lies in not including KUs from the third-party libraries that are used by the studied projects. Our aim was to gauge the effectiveness of KUs to reflect the expertise of developers, leading us to concentrate solely on KUs developed by the developers of studied projects. To minimize the risk of overlooking KUs that are implemented within these projects, we meticulously applied type binding resolutions using the Java JDT library~\citep{eclipse_jdt}.

A potential threat in our study arises from the need to accurately link GitHub accounts with commit author names, as they often differ. To mitigate this threat and prevent bias, we employ a conservative selection process for developers, only including those where we can confidently match their commit author names with their GitHub display names. In doing so, we prioritize precision over recall to ensure the integrity of our analysis.


\smallskip \noindent \textbf{External validity.} In this study, we only focus on Java projects of GitHub. We were aware of the fact that GitHub contains projects that are not real-world projects (e.g., academic projects and toy projects). Therefore, we carefully filtered our studied projects using the list of real-world software projects curated by~\citet{engineered_project_github_EMSE_2017}. We encourage future studies to broaden the scope of our study and investigate how our findings apply to software projects written in other popular programming languages (e.g., Python) that have certification exams in different development areas (e.g., web development). For example, popular programming languages for web development (e.g., Ruby on Rails, JavaScript, Python, ASP.Net, Node-js, and Objective-C) have their own certification exams \citep{web_certification_exam}.

One other threat to the validity of our study is our reliance on the GHTorrent dataset, which ceased updating its GitHub data collection. To mitigate this issue, we specifically collected the last available snapshot from GHTorrent dated March 06, 2021. By using this most recent snapshot, we aim to ensure that our dataset includes the latest available comprehensive data from GitHub, thereby minimizing the impact of any discontinuation and maximizing the relevance and accuracy of our analysis.

    \section{Conclusion}
\label{sec:Conclusion}
\sloppy This paper presents an empirical study on how to leverage KUs of programming language to predict long time contributors (LTCs). We carefully engineer KU-based features across five distinct dimensions and construct a classification model named KULTC for predicting LTCs. Our findings show that KULTC outperforms a state-of-the-art's baseline model (BAOLTC), achieving an AUC of at least 0.75. In addition, combining features of BAOLTC with KULTC results in an even higher-performing model (KULTC+BAOLTC) that outperforms individual KULTC and BAOLTC. The normalized AUC improvement of KULTC+BAOLTC over BAOLTC is 14.5\%$\sim$16.5\% across different settings for LTC prediction. We also show that a more cost-effective model, which is built with single feature dimension of KULTC and all features of BAOLTC, outperforms BAOLTC. Our model interpretation analysis reveals that KU-based developer expertise in the studied projects is the most influential feature dimension for predicting LTC. Even though we cannot claim causation, our empirical evidence together with prior literature reinforce the role of programming language expertise in predicting LTCs.

Our encouraging results highlight the importance of programming language skills in predicting LTCs. Future studies should explore richer conceptualization and operationalization of KUs (e.g., supporting the detection of architectural design concepts of projects) and evaluate whether our findings generalize to other projects across different programming languages.

	
	\section*{Data Availability Statement (DAS)}
\label{sec:Data_Availability_Statement}

A supplementary material package is provided online in the following link:
\url{https://shorturl.at/dpKU4}. The contents will be made available on a public GitHub
repository once the paper is accepted.
	\section*{Funding and/or Conflicts of interests/Competing interests}
\label{sec:Conflict_Of_Interest}
The authors declared that they have no conflict of interest.
	
	\begin{footnotesize}
		\bibliographystyle{spbasic}      
		\bibliography{bib/references.bib}   
	\end{footnotesize}	

	\clearpage
	
	\appendix

	\begin{Large}
		\noindent \textbf{Appendix}
	\end{Large}
	
	\normalsize
	\vspace{-1ex}
	\section{Java Certification Exams and Knowledge Units}
\label{appendix:cert-exams}
\vspace{-1cm}
\begin{table*}[!b]
    \centering
    \caption{Knowledge Units derived from the Java SE 8 Programmer I Exam, Java SE 8 Programmer II Exam, and Java EE Developer Exam}
    \label{tab:ku-from-java-exams}
    \resizebox{\textwidth}{!}{
    \begin{tabular}{p{3.0cm}p{16cm}}
    \toprule
    \multicolumn{1}{C{3.0cm}}{\textbf{Knowledge Unit (KU)}} & \multicolumn{1}{C{16cm}}{\textbf{Key Capabilities}} \\ \midrule
    
    \textbf{[K1]} Data Type & 
    \textbf{[C1]} Declare and initialize different types of variableS(e.g., primitive type, parameterized type, and array type), including the casting of primitive data types
    \\ \midrule
    \textbf{[K2]} Operator and Decision &
    \textbf{[C1]} Use Java operators(e.g., assignment and postfix operators); use parentheses to override operator precedence \newline
    \textbf{[C2]} Test equality between strings and other objects using \texttt{==} and \texttt{equals()} \newline
    \textbf{[C3]} Create and use \texttt{if}, \texttt{if-else}, and ternary constructs \newline
    \textbf{[C4]} Use a \texttt{switch} statement
    \\ \midrule
    
    \textbf{[K3]} Array &
    \textbf{[C1]} Declare, instantiate, initialize and use a one-dimensional array \newline
    \textbf{[C2]} Declare, instantiate, initialize and use a multi-dimensional array 
    \\ \midrule
    
    \textbf{[K4]} Loop &
    \textbf{[C1]} Create and use \texttt{while} loops \newline
    \textbf{[C2]} Create and use \texttt{for} loops, including the \texttt{enhanced for} loop \newline
    \textbf{[C3]} Create and use \texttt{do-while} loops \newline
    \textbf{[C4]} Use \texttt{break} statement \newline
    \textbf{[C5]} Use \texttt{continue} statement
    \\ \midrule
    
    \textbf{[K5]} Method and Encapsulation &
    \textbf{[C1]} Create methods with arguments and return values \newline
    \textbf{[C2]} Apply the ``static'' keyword to methods, fields, and blocks \newline
    \textbf{[C3]} Create an overloaded method and overloaded constructor \newline
    \textbf{[C4]} Create a constructor chaining (use ``this()'' method to call one constructor from another constructor \newline
    \textbf{[C5]} Use variable length arguments in the methods \newline
    \textbf{[C6]} Use different access modifiers (e.g., private and protected) other than ``default'' \newline
    \textbf{[C7]} Apply encapsulation: identify set and get method to initialize any private class variables \newline
    \textbf{[C8]} Apply encapsulation: Immutable class generation-final class and initialize private variables through the constructor
    \\ \midrule
    
    \textbf{[K6]} Inheritance &
    \textbf{[C1]} Use basic polymorphism (e.g., a superclass refers to a subclass) \newline
    \textbf{[C2] }Use polymorphic parameter (e.g., pass instances of a subclass or interface to a method) \newline
    \textbf{[C3]} Create overridden methods \newline
    \textbf{[C4]} Create ``abstract'' classes and ``abstract'' methods \newline
    \textbf{[C5]} Create ``interface'' and implement the interface \newline
    \textbf{[C6]} Use ``super()'' and the ``super'' keyword to access the members(e.g., fields and methods) of a parent class \newline
    \textbf{[C7]} Use casting in referring a subclass object to a superclass object
    \\ \midrule
    
    \textbf{[K7]} Advanced Class Design &
    \textbf{[C1]} Create inner classes, including static inner classes, local classes, nested classes, and anonymous inner classes \newline
    \textbf{[C2]} Develop code that uses the final \newline
    \textbf{[C3]} Use enumerated types including methods and constructors in an ``enum'' type \newline
    \textbf{[C4]} Create singleton classes and immutable classes
    \\ \midrule 
    
    \textbf{[K8]} Generics and Collection &
    \textbf{[C1]} Create and use a generic class \newline
    \textbf{[C2]} Create and use \texttt{ArrayList}, \texttt{TreeSet}, \texttt{TreeMap}, and \texttt{ArrayDeque} \newline
    \textbf{[C3]} Use \texttt{java.util.Comparator} and \texttt{java.lang.Comparable} interfaces \newline
    \textbf{[C4]} Iterate using forEach methods of List
    \\ \midrule

    \textbf{[K9]} Functional Interface &
    \textbf{[C1]} Use the built-in interfaces included in the \texttt{java.util.function} packages such as \texttt{Predicate}, \texttt{Consumer}, \texttt{Function}, and \texttt{Supplier} \newline
    \textbf{[C2]} Develop code that uses primitive versions of functional interfaces \newline
    \textbf{[C3]} Develop code that uses binary versions of functional interfaces \newline
    \textbf{[C4]} Develop code that uses the \texttt{UnaryOperator} interface
    \\ \midrule
    
    \textbf{[K10]} Stream API &
    \textbf{[C1]} Develop code to extract data from an object using \texttt{peek()} and \texttt{map()} methods, including primitive versions of the \texttt{map()} method \newline
    \textbf{[C2]} Search for data by using search methods of the Stream classes, including \texttt{findFirst}, \texttt{findAny}, \texttt{anyMatch}, \texttt{allMatch}, \texttt{noneMatch} \newline
    \textbf{[C3]} Develop code that uses the Optional class \newline
    \textbf{[C4]} Develop code that uses Stream data methods and calculation methods \newline
    \textbf{[C5]} Sort a collection using Stream API \newline
    \textbf{[C6]} Save results to a collection using the collect method \newline
    \textbf{[C7]} \texttt{UseflatMap()} methods in the Stream API
    \\ \midrule
    
    \textbf{[K11]} Exception &
    \textbf{[C1]} Create a try-catch block \newline
    \textbf{[C2]} Use catch, multi-catch, and finally clauses \newline
    \textbf{[C3]} Use autoclose resources with a try-with-resources statement \newline
    \textbf{[C4]} Create custom exceptions and autocloseable resources \newline
    \textbf{[C5]} Create and invoke a method that throws an exception \newline
    \textbf{[C6]} Use common exception classes and categories(such as \texttt{NullPointerException}, \texttt{ArithmeticException}, \texttt{ArrayIndexOutOfBoundsException}, \texttt{ClassCastException}) \newline
    \textbf{[C5]} Use assertions
    \\ \midrule
    
    \textbf{[K12]} Date Time API &
    \textbf{[C1]} Create and manage date-based and time-based events including a combination of date and time into a single object using \texttt{LocalDate}, \texttt{LocalTime}, \texttt{LocalDateTime}, \texttt{Instant}, \texttt{Period}, and \texttt{Duration} \newline
    \textbf{[C2]} Formatting date and times values for using different timezones \newline
    \textbf{[C3]} Create and manage date-based and time-based events using Instant, Period, Duration, and Temporal Unit \newline
    \textbf{[C4]} Create and manipulate calendar data using classes from \texttt{java.time.LocalDateTime}, \texttt{java.time.LocalDate}, \texttt{java.time.LocalTime}, \texttt{java.time.format.DateTimeFormatter}, and \texttt{java.time.Period} 
    \\ \midrule
    
    \textbf{[K13]} IO &
    \textbf{[C1]} Read and write data using the console \newline
    \textbf{[C2]} Use \texttt{BufferedReader}, \texttt{BufferedWriter}, \texttt{File}, \texttt{FileReader}, \texttt{FileWriter}, \texttt{FileInputStream}, \texttt{FileOutputStream}, \texttt{ObjectOutputStream}, \texttt{ObjectInputStream}, and \texttt{PrintWriter} in the \texttt{java.io} package
    \\ \midrule
    
    \textbf{[K14]} NIO &
    \textbf{[C1]} Use the Path interface to operate on file and directory paths \newline
    \textbf{[C2]} Use the Files class to check, read, delete, copy, move, and manage metadata a file or directory
    \\ \midrule
    
    \textbf{[K15]} String Processing &
    \textbf{[C1]} Search, parse and build strings \newline
    \textbf{[C2]} Manipulate data using the \texttt{StringBuilder} class and its methods \newline
    \textbf{[C3]} Use regular expression using the \texttt{Pattern} and \texttt{Matcher} class \newline
    \textbf{[C4]} Use string formatting
    \\ \midrule
    \end{tabular}
       }
\end{table*}

\begin{table*}[!htbp]
    \centering
    \resizebox{\textwidth}{!}{
    \begin{tabular}{p{4.0cm}p{15cm}}
    \toprule
    \multicolumn{1}{C{3.0cm}}{\textbf{Knowledge Unit (KU)}} & \multicolumn{1}{C{11cm}}{\textbf{Key Capabilities}} \\ \midrule
    
    \textbf{[K16]} Concurrency &
    \textbf{[C1]} Create worker threads using \texttt{Runnable}, \texttt{Callable} and use an \texttt{ExecutorService} to concurrently execute tasks
    \textbf{[C2]} Use \texttt{synchronized} keyword and \texttt{java.util.concurrent.atomic} package to control the order of thread execution \newline
    \textbf{[C3]} Use \texttt{java.util.concurrent}   collections and classes including \texttt{CyclicBarrier} and \texttt{CopyOnWriteArrayList} \newline
    \textbf{[C4]} Use parallel Fork/Join Framework
    \\ \midrule
    
    \textbf{[K17]} Database &
    \textbf{[C1]} Describe the interfaces that make up the core of the JDBC API, including the \texttt{Driver}, \texttt{Connection}, \texttt{Statement}, and \texttt{ResultSet} interfaces \newline
    \textbf{[C2]} Submit queries and read results from the database, including creating statements, returning result sets, iterating through the results, and properly closing result sets, statements, and connections
    \\ \midrule
    \textbf{[K18]} Localization &
    \textbf{[C1]} Read and set the locale by using the Locale object \newline
    \textbf{[C2]} Build a resource bundle for each locale and load a resource bundle in an application
    \\ \midrule
    \textbf{[K19]} Java Persistence &
    \textbf{[C1]} Create JPA Entity and Object-Relational Mappings (ORM) \newline
    \textbf{[C2]} Use Entity Manager to perform database operations, transactions, and locking with JPA entities \newline
    \textbf{[C3]} Create and execute JPQL statements \\ \midrule
    
    \textbf{[K20]} Enterprise Java Bean &
    \textbf{[S1]} Create session EJB components containing synchronous and asynchronous business methods, manage the life cycle container callbacks, and use interceptors. \newline
    \textbf{[S2]} Create EJB timers \\ \midrule

    \textbf{[K21]} Java Message Service API &
    \textbf{[S1]} Implement Java EE message producers and consumers, including Message-Driven beans \newline
    \textbf{[S2]} Use transactions with JMS API \\ \midrule
    
    \textbf{[K22]} SOAP Web Service &
    \textbf{[S1]} Create SOAP Web Services and Clients using JAX-WS API  \newline
    \textbf{[S2]} Create marshall and unmarshall Java Objects by using JAXB API \\ \midrule

    \textbf{[K23]} Servlet &
    \textbf{[S1]} Create Java Servlet and use HTTP methods \newline
    \textbf{[S2]} Handle HTTP headers, parameters, cookies \newline
    \textbf{[S3]} Manage servlet life cycle with container callback methods and WebFilters \\ \midrule

    \textbf{[K24]} Java REST API &
    \textbf{[S1]} Apply REST service conventions \newline
    \textbf{[S2]} Create REST Services and clients using JAX-RS API \\ \midrule

    \textbf{[K25]} Websocket &
    \textbf{[S1]} Create WebSocket Server and Client Endpoint Handlers \newline
    \textbf{[S3]} Produce and consume, encode and decode WebSocket messages \\ \midrule

    \textbf{[K26]} Java Server Faces &
    \textbf{[S1]} Use JSF syntax and use JSF Tag Libraries \newline
    \textbf{[S2]} Handle localization and produce messages \newline
    \textbf{[S3]} Use Expression Language (EL) and interact with CDI beans \\ \midrule
    
    \textbf{[K27]} Contexts and Dependency Injection (CDI) &
    \textbf{[S1]} Create CDI Bean Qualifiers, Producers, Disposers, Interceptors, Events, and Stereotypes \\ \midrule 

    \textbf{[K28]} Batch Processing &
    \textbf{[S1]} Implement batch jobs using JSR 352 API \\
    \bottomrule
    \end{tabular}
   }
\end{table*}

\clearpage

\section{BAOLTC Model}

In the following, we describe the five dimensions from which model features were engineered. Table~\ref{tab:bao_model_feature} lists those features and provides a brief explanation for each.

\begin{description}[wide = 0pt, itemsep = 3pt, topsep=3pt]
    \item[\textbf{Developer profile.}] This dimension refers to the features that are extracted from a newcomer's information when the newcomer submits his/her initial commit to a project. This dimension includes a total of eight features. These features are dependent on the historical activities of a newcomer. For example, \textit{user-own-repos} quantifies the number of projects owned by the new developers. The \textit{user-history commits}, \textit{user-history-issues} and \textit{user-history-pull-requests} features measure the previous activities of the new developers in GitHub.
    \item[\textbf{Repository profile.}] This dimension captures the historical activities of all contributors in a project when a newcomer submits his/her first commit to the project. For example, \textit{before-repo-commits} quantifies the number of commits that the repository have when the new developers join the project. This dimension contains a total of 20 different features that are extracted from six sources of information of the project: programming language, commits, contributors, issues, pull-requests and watchers.
    \item[\textbf{Developer monthly activity.}] This dimension refers to the features that are extracted from the first month's activities of a newcomer since the newcomer submits his/her first commit. This dimension contains 12 different features that are related to the newcomer's initial activity in the project. For example, \textit{month-user-commits} counts the number of commits that the new developer submits to the project in his/her first month.
    \item[\textbf{Repository monthly activity.}] This dimension focuses on features that are extracted from the activities of all contributors in the first month after a newcomer joins a project. It encompasses 18 distinct features. For instance, \textit{month-repo-commit} measures the total commits made to the repository in the newcomer's first month, while \textit{month-repo-pull-requests} count the total number of pull requests submitted to the repository in the newcomer's first month.
    \item[\textbf{Collaboration Network.}] This dimension consists of the features that are based on collaboration activities between a newcomer and other contributors in the first month after the newcomer joins the project. These features are extracted from a graph that is built based on authors of commits, issues and PRs and the developers who provide comments on those commits, issues and PRs. A total of five social network features are extracted from this graph that are used to build a prediction model (e.g., \textit{degree-centrality}, \textit{closeness-centrality}, \textit{betweenness-centrality}, \textit{eigenvector-centrality} and \textit{clustering-coefficient}).
\end{description}

\begin{table}[!htbp]
    \centering
    \caption{The list of features that are used in the study of \citet{bao2019large}.}
    \label{tab:bao_model_feature}
    \resizebox{\columnwidth}{!}{
    \begin{tabular}{p{2cm}p{5.3cm}p{15cm}}
        \toprule
        \multicolumn{1}{C{2cm}}{\textbf{Dimension}} & 
        \multicolumn{1}{C{5.3cm}}{\textbf{Factor}} &
        \multicolumn{1}{C{15cm}}{\textbf{Explanation}}  \\ \midrule

        \multirow{8}{2cm}{Developer Profile}  & user-age &  Number of days between the registration date of the new developer and the date that he joins the repository\\
         & user-own-repos & Number of repositories the new developer owns when he joins the repository \\
        & user-watch-repos & Number of repositories the new developer watches when he joins the repository \\
        & user-contribute-repos & Number of repositories in which the new developer has submitted at least one commit when he joins the repository \\
        & user-history-commits & Number of commits the new developer has submitted when he joins the repository \\
        & user-history-pull-requests & Number of pull requests the new developer has submitted when he joins the repository
        user \\
        & user-history-issues & Number of issues the new developer has submitted when he joins the repository \\
        & user-history-followers & Number of users who follow the new developer
        language \\ \midrule

        \multirow{14}{2cm}{Repository Profile}  & language & Main programming language used by the repository before \\
        &before-repo-commits & Number of commits that the repository have when the new developer joins \\
        & before-repo-commit-comments & Number of commit comments that the repository has when the new developer joins \\
        & before-repo-contributors & Number of contributors that the repository has when the new developer joins \\
        & before-repo-contributor{S} & Statistics of commits of contributors in the repository, where S can be max, min, mean, median, and std \\ 
        & before-repo-issues & Number of issues that the repository have when the new developer joins \\ 
        & before-repo-issue-comments & Number of issue comments that the repository has when the new developer joins \\
        & before-repo-issue-events & Number of issue events that the repository has when the new developer joins \\
        & before-repo-issue-events-closed & Number of issue closed events that the repository has when the new developer joins \\
        & before-repo-issue-events-assigned & Number of issue assigned events that the repository has when the new developer joins \\
        & before-repo-pull-requests & Number of pull requests that the repository has when the new developer joins \\
        & before-repo-pull-request-comments & Number of pull request comments that the repository has when the new developer joins \\
        & before-repo-pull-request-history & Number of pull request events the repository has when the new developer joins \\
        & before-repo-pull-request-history-merged & Number of merged pull request events the repository has when the new developer joins \\ 
        & before-repo-pull-request-history-closed & Number of closed pull request events the repository has when the new developer joins \\ 
        & before repo watchers & Number of watchers the repository has when the new developer joins \\ \midrule

        \multirow{10}{2cm}{Developer  Monthly Activity}  & month-user-commits & Number of commits that the new developer submits to the repository in the first month \\
        & month-user-commit-comments & Number of comments received in the commits submitted by the new developer in the first month \\
        & month-user-issues & Number of issues that the new developer submits to the repository in the first month \\
        & month-user-issue-comments & Number of comments received in the issues submitted by the new developer in the first month \\
        & month-user-issue-events & Number of events received in the issues that the new developer submits to the repository in the first month \\
        & month-user-issue-events-closed & Number of closed events received in the issues submitted by the new developer in the first month \\
        & month-user-issue-events-assigned & Number of assigned events received in the issues submitted by the new developer in the first month \\
        & month-user-pull-requests & Number of pull request that the new developer submits to the repository in the first month \\
        & month-user-pull-request-comments & Number of comments received in the pull request submitted by the new developer in the first month \\
        & month-user-pull-request-history & Number of pull request events received in the pull request submitted by the new developer in the first month \\
        & month-user-pull-request-history-merged &  Number of merged pull request events received in the pull request submitted by the new developer in the first month \\
        & month-user-pull-request-history-closed & Number of closed pull request events received in the pull request submitted by the new developer in the first month \\ \midrule

        \multirow{10}{2cm}{Repository  Monthly Activity} & month-repo-commits & Number of commits submitted to the repository in the first month that the new developer joins \\
        & month-repo-commit-comments & Number of comments received in the commits submitted to the repository in the first month \\ 
        & month-repo-contributors & Number of contributors who submitted at least one commits to the repository in the first month \\
        & month-repo-contributor{S} & Statistics of commits of contributors in the first month,where S can be max, min, mean, median, and std \\ 
        & month-repo-issues & Number of issues submitted to the repository in the first month \\
        & month-repo-issue-comments  & Number of comments received in the issues submitted to the repository in the first month \\ 
        & month-repo-issue-events & Number of events received in the issues submitted to the repository in the first month \\ 
        & month-repo-issue-events-closed & Number of closed events received in the issues submitted to the repository in the first month \\
        & month-repo-issue-events-assigned & Number of assigned events received in the issues submitted to the repository in the first month \\
        & month-repo-pull-requests & Number of pull requests submitted to the repository in the first month \\ 
        & month-repo-pull-request-comments & Number of comments received in the pull requests submitted to the repository in the first month \\ 
        & month-repo-pull-request-history & Number of events received in the pull requests submitted to the repository in the first month \\ 
        & month-repo-pull-request-history-merged & Number of merged events received in the pull requests submitted to the repository in the first month \\
        & month- pull-request-history-closed & Number of closed events received in the pull requests submitted to the repository in the first month \\ \midrule

        \multirow{4}{2cm}{Collaboration  Network} & degree-centrality & \multirow{4}{12cm}{These metrics are used to quantify a newcomer’s degree of activity in the collaboration structure of an OSS project} \\
        & closeness-centrality & \\
        & betweenness-centrality  & \\
        & eigenvector-centrality-clustering & \\ \bottomrule

    \end{tabular}
    }
\end{table}

\end{document}